\documentclass[fleqn,usenatbib, useAMS]{mnras}

\usepackage{newtxtext,newtxmath}
\usepackage[T1]{fontenc}
\usepackage{ae,aecompl}
\usepackage[utf8]{inputenc}
\usepackage{threeparttable}
\usepackage{booktabs, caption, makecell}
\usepackage{pdflscape}
\usepackage{geometry}
\usepackage{float}
\usepackage{pifont}
\usepackage{orcidlink}

\usepackage{graphicx}	
\usepackage{amsmath}	

\newcommand{\Line}[3]{\Ion{#1}{#2}~#3\,\AA}

\newcommand{\Ion}[2]{#1{\,\sc#2}}

\usepackage{rotating}
\usepackage{float}
\usepackage{tabularx}
\usepackage{natbib}
\usepackage{hyperref}
\usepackage{academicons}
\newcommand{\Teff}{\mbox{$T_{\mathrm{eff}}$}}
\newcommand{\logg}{\mbox{$\log g$}}
\newcommand{\Msun}{\mbox{$\mathrm{M_\odot}$}}
\newcommand{\Rsun}{\mbox{$\mathrm{R_\odot}$}}
\newcommand{\gs}{\mbox{$\mathrm{g\,s^{-1}}$}}
\newcommand{\kms}{\mbox{$\mathrm{km\,s^{-1}}$}}

\newcommand{\psj}{Planetary Science}

\title[Characterisation of an icy exo-planetesimal]{Discovery of an icy and nitrogen-rich extrasolar planetesimal}
 
\author[Sahu et al.]
{Snehalata Sahu$^{\orcidlink{0000-0002-0801-8745}}$,$^{1}$\thanks{E-mail: snehalatash30@gmail.com}
Boris~T.~G\"ansicke$^{\orcidlink{0000-0002-2761-3005}}$,$^{1,2}$ Jamie T. Williams$^{\orcidlink{0009-0007-8709-9689}}$,$^{1}$ Detlev G. Koester$^{\orcidlink{0000-0002-6164-6978}}$,$^{3}$ Jay Farihi$^{\orcidlink{0000-0003-1748-602X}}$,$^{4}$ \newauthor Steven J. Desch$^{\orcidlink{0000-0002-1571-0836
}}$,$^{5}$ Nicola Pietro Gentile Fusillo $^{\orcidlink{0000-0002-6428-4378}}$,$^{6,7}$ Dimitri Veras$^{\orcidlink{0000-0001-8014-6162}}$,$^{1,8,9}$ Sean N. Raymond$^{\orcidlink{0000-0001-8974-0758}}$,$^{10}$ \newauthor Maria Teresa Belmonte$^{\orcidlink{0000-0002-3527-201X}}$~$^{11}$ \\\\
$^{1}$ Department of Physics, University of Warwick, Coventry, CV4 7AL, UK\\
$^{2}$ Instituto de Astrofísica de Canarias, E-38205 La Laguna, Tenerife, Spain\\
$^{3}$ Institut f\"ur Theoretische Physik und Astrophysik, Christian-Albrechts-Universit\"at, 24118 Kiel, Germany\\
$^{4}$ Department of Physics and Astronomy, University College London, London, WC1E 6BT, UK\\
$^{5}$ School of Earth and Space Exploration, Arizona State University, PO Box 876004, Tempe AZ USA 85287-6004\\
$^{6}$ Department of Physics, Universit\`a degli Studi di Trieste, Via A. Valerio 2, I-34127, Trieste, Italy\\
$^{7}$ INAF – Osservatorio Astronomico di Trieste, via Tiepolo 11, I-34131, Trieste, Italy\\
$^{8}$ Centre for Exoplanets and Habitability, University of Warwick, Coventry, CV4 7AL, UK\\
$^{9}$ Centre for Space Domain Awareness, University of Warwick, Coventry, CV4 7AL, UK\\
$^{10}$ Laboratoire d’Astrophysique de Bordeaux, CNRS and Universit{\'e} de Bordeaux, All{\' e}e Geoffroy St. Hilaire, F-33165 Pessac, France\\
$^{11}$ Universidad de Valladolid, Departamento de F\'isica Te\'orica, At\'omica y \'Optica, P.$^\circ$ de Bel\'en, 7, Valladolid 47011, Spain
}

\begin{document}
\label{firstpage}
\pagerange{\pageref{firstpage}--\pageref{lastpage}}

\maketitle

\begin{abstract}
White dwarfs accreting planetary debris provide detailed insight into the bulk composition of rocky exo-planetesimals. However, only one Kuiper-Belt analogue has been identified in that way so far. 
Here, we report the accretion of an icy extra-solar planetesimal onto white dwarf WD\,1647+375 using ultraviolet spectroscopy from the \textit{Hubble Space Telescope}. The accreted material is rich in the volatiles carbon, nitrogen, and sulphur, with a chemical composition analogous to Kuiper-belt objects (KBOs) in our solar system. It has a high nitrogen mass fraction ($5.1\pm1.6$\,per cent) and large oxygen excess ($84\pm7$\,per cent), indicating that the accreted planetesimal is water-rich (a water-to-rock ratio of $\simeq2.45$), corroborating a cometary- or dwarf planet-like composition. The white dwarf has been accreting at a rate of $\approx2\times10^{8}\,\gs$ for the past 13 years, implying a minimum mass of $\sim10^{17}$\,g for the icy parent body. The actual mass could be several orders of magnitude larger if the accretion phase lasts $\sim10^5$\,yr as estimated in the literature from debris disc studies. We argue that the accreted body is most likely a fragment of a KBO dwarf planet based on its nitrogen-rich composition. However, based on the chemical composition alone, it is difficult to discern whether this icy body is intrinsic to this planetary system, or may have an interstellar origin.
\end{abstract}

\section{Introduction}
The solar system contains a vast number of icy bodies, including comets and the moons of giant planets. These bodies originate in distant regions of the solar system, such as the Kuiper-belt and the Oort cloud, where temperatures are sufficiently low for volatile ices to remain in solid form. These objects are predominantly composed of water, CO$_{2}$, and NH$_{3}$, and are representative of the primitive and unaltered matter from the early solar nebula. They offer key insights into the composition of protoplanetary discs and thus are important for understanding the delivery of water (and potentially the origins of life \citealt{anders89-1, anslowetal23-1}) onto terrestrial planets. The source of water on Earth remains an intensively debated topic (see e.g., \citealt{pianietal20-1, kraletal24-1,barrettetal25-1}), where a few recent studies suggest that water could be produced by the chemical reactions of primordial hydrogen-dominated atmospheres with magma oceans \citep{young2023, Rogers2025}.

Beyond the solar system, a small number of exo-comets (see the review by \citealt{strom2020}) have been discovered via high-resolution spectra of transient absorption features (such as \ion{Ca}{II}, \ion{Fe}{II}) caused by the dusty tails of comets while transiting the star \citep{keifer2014, keifer2014b, welsh2018}. These observations are challenging, as these icy exo-planetary bodies are small and faint. Since typically only gases are detected, our insight into the detailed chemical composition of the exo-comets remains limited. In this respect, white dwarfs have the potential to serve as powerful diagnostic tools for investigating the bulk composition of exo-planetary icy bodies.

White dwarfs with $\Teff<25\,000$\,K typically have simple atmospheres dominated by hydrogen or helium, as heavy elements formed during earlier evolutionary stages sink to the core under the influence of their high surface gravities \citep{Evry1945}. The sinking timescales for these metals are always short (ranging from days to millions of years, depending on whether the atmosphere is dominated by H or He, (see \citealt{koester2009}) compared to the time the star has spent in the white dwarf phase. The detection of metals in their atmospheres is therefore attributed to external sources, primarily from the accretion of planetary debris \citep{Jura2003}, as confirmed by many studies in the past two decades \citep{klein2010, farihi2010, kawka2012, hollands2017, pi2021}. This phenomenon makes them powerful probes of the bulk composition of exo-planetary bodies \citep{zuckerman2007}.

Optical and ultraviolet spectroscopic surveys have shown that between 25\,per cent \citep{zuckerman2003} to 50\,per cent \citep{koester2014} of white dwarfs exhibit photospheric metals. Among the heavy elements, O, Mg, Si, Ca, and Fe are typically the most abundant, consistent with the accretion of rocky bodies. Detailed analyses of a small number of heavily metal-enriched white dwarfs have revealed abundance patterns that closely resemble those of bulk Earth or CI chondrites \citep{gansicke2012, dufour2017, doyleetal23-1}. In contrast, only a handful of systems have been found accreting volatile-rich material containing elements such as C, N, S or excess oxygen that is indicative of water-rich bodies \citep{farihi2013, raddi2015, xu2017, GF2017}. 

Given the ubiquity of icy bodies in our solar system, the observational findings discussed above raise an important question: Why is the detection of white dwarfs accreting volatile-rich material so rare? Whereas comets and Kuiper-belt-like objects are numerous in the Solar system, and very likely also in exo-planetary systems, a large fraction of them are likely to be ejected during the metamorphosis of their host stars into white dwarfs \citep{verasetal14-3, stoneetal15-1}. In addition, there is a strong observational selection effect: cool white dwarfs ($<13\,000$\,K) usually dominate the spectroscopic surveys which are less sensitive to detecting volatiles (such as C, N, O) due to their weaker transitions in optical wavelengths \citep{jamie2024}. Furthermore, recent studies have highlighted other possible explanations, including observational biases due to asynchronous accretion \citep{malamud2016, brouwers2023}, and the shielding effect of white dwarf magnetospheres, which can prevent the accretion of volatile vapours (produced by sublimating comet fragments) within the co-rotation radius \citep{zhou2024}. 

To identify and characterise the volatile-enriched white dwarf systems, measuring the abundance of nitrogen is crucial as it serves as a key indicator of exo-planetary ices or primitive material enriched with volatile elements, similar to those found in comets, interstellar objects \citep[Oumuamua][]{jackson2021}, Kuiper-Belt objects, and the crusts of icy moons of the outer solar system. The detection of nitrogen requires far-ultraviolet spectroscopy, and has so far only been reported for three of the white dwarfs with He \citep{xu2017, klein2021} and H-rich atmospheres \citep{johnson2022}. The overall chemical composition of the debris accreted by these three white dwarfs indicates that they accreted Kuiper belt-like objects. 

Here, we report nitrogen detection in a warm (22\,040\,K) hydrogen-rich atmosphere of WD\,1647+375 using far-ultraviolet spectroscopy obtained with \textit{Hubble Space Telescope (HST)}. In addition to nitrogen, we detect other volatile elements (C, O, S) which, combined with its large oxygen excess, imply that this white dwarf is accreting an icy exo-planetesimal.

\section{Observation and Data Analysis}\label{sec:analysis}
We obtained ultraviolet and optical spectroscopy to measure the metal abundances in the atmosphere of WD\,1647+375. Below, we provide details on these observations and describe the fitting technique as well as the computation of the accretion rates calculated from the best-fit abundances.

\subsection{Ultraviolet spectroscopy}
WD\,1647+375 was observed twice with the Cosmic Origins Spectrograph (COS) onboard \textit{HST}.  An initial observation with an exposure time of 780\,s was obtained on 2011 October 5 as part of the snapshot program 12474, followed by a deep exposure using a 2164\,s integration on 2012 December 16 under program 12869 (PI Boris G\"ansicke for both programs). On both occasions, we used the G130M grating with a central wavelength of 1291\,\AA\ which covers the wavelength range 1130$-$1435\,\AA, with a gap at 1278$-$1288\,\AA\ due the space between two detector segments. The resolving power of the grating is $R=16\,000$. The spectra were processed with the COS pipeline CALCOS\,3.1.8. 

The snapshot spectrum was acquired with the lifetime position LP1, and the deeper exposure spectrum at LP2. Since the COS wavelength accuracy corresponds to a velocity of $15\,\kms$, we used the spectra of WD\,1647+375 to check for differences between the measurements arising from the wavelength calibration in different LP settings. Comparing line velocities, we find that there is an average velocity offset of $\approx10-12$\,\kms\ between the \Line{N}{i}{1243} doublet and other strong absorption lines of C, Si and O within the deep exposure.  In contrast, all line velocities are consistent in the case of snapshot spectrum (Table\,\ref{tab:vel_comp}). 
Inspecting the two COS spectra, we find a relative shift in the wavelength calibration of deep exposure ($\approx0.04$\,\AA), particularly in the 1200$-$1300\,\AA\ spectral region. This velocity difference may arise due to an error in COS wavelength calibration, which depends on wavelength accuracy, geometric distortions, and drifts \citep{olive2010}.

\begin{table}
\caption{Comparison of line velocities derived from two COS observing programs of WD\,1647+375 with exposure times of 780\,s (snapshot) and 2164\,s (deep).}
\begin{center}
\begin{tabular}{lccc}
\hline
Metal	&	Wavelength	region	(\AA)	&	\multicolumn{2}{c}{V$_{r}$ (\kms)}	\\
&	&	snapshot	&	deep	\\\hline	
\ion{C}{iii}	&	1175$-$1177	&	14.37	$\pm$	0.55	&	13.04	$\pm$	0.36	\\
\ion{C}{ii}*	&	1334$-$1335	&	12.24	$\pm$	0.38	&	13.74	$\pm$	0.26	\\
\ion{N}{i}*	&	1199$-$1200	&	14.84	$\pm$	6.18	&	7.18	$\pm$	2.75	\\
\ion{N}{i}	&	1242$-$1245	&	15.29	$\pm$	3.59	&	1.42	$\pm$	2.81	\\
\ion{O}{i}	&	1151$-$1153	&	18.40	$\pm$	0.93	&	13.01	$\pm$	0.64	\\
\ion{O}{i}*	&	1302$-$1306	&	19.75	$\pm$	0.38	&	19.1	$\pm$	0.23	\\
\ion{Al}{iii}	&	1379$-$1384	&	12.67	$\pm$	1.05	&	12.39	$\pm$	0.69	\\
\ion{Si}{ii}	&	1260$-$1265	&	18.41	$\pm$	0.29	&	10.95	$\pm$	0.19	\\
\ion{Si}{iii}*	&	1298$-$1301	&	20.10	$\pm$	0.26	&	17.92	$\pm$	0.18	\\

\hline
\end{tabular}
\end{center}
\footnotesize 
Note: *these photospheric features are contaminated by lines from the interstellar medium (ISM)
\label{tab:vel_comp}
\end{table} 

\begin{figure}
    \centering   
    \includegraphics[width=\columnwidth]{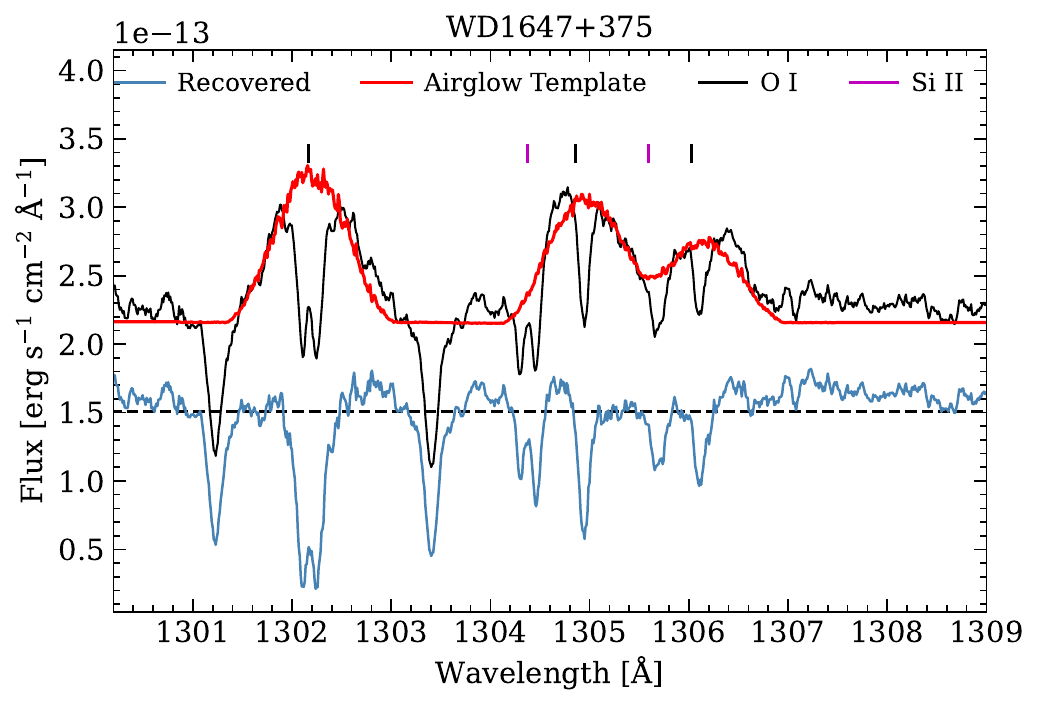}
    \caption{Airglow correction of WD\,1647+375 of the deep exposure where the raw spectrum and the best-fit airglow template are shown in black and red, respectively. The airglow-corrected spectrum is shown in blue and arbitrarily shifted in flux for clarity. The strong absorption lines of \ion{Si}{ii} and \ion{O}{i} affected by the airglow are marked in the plot. The dashed black line shows the average continuum flux in the wavelength region 1300$-$1309\,\AA.}
    \label{fig:ag_corr}
\end{figure}

The \ion{O}{i} triplet (1302$-$1306\,\AA) in the spectra can be affected by the geocoronal lines whose intensity varies as a function of \textit{HST}'s orbital position (i.e. the observatory's ``day'' and ``night'' time). Both \textit{HST} spectra of WD\,1647+675 are strongly affected by airglow. We used community-generated templates for COS \citep{bourier2018} to remove this feature from the spectra \citep{Williams2025} as shown in Fig.\,\ref{fig:ag_corr}. The corrected spectra for both exposures were used for determining the oxygen abundance. 

\subsection{Optical spectroscopy}
A low resolution ($R\approx1000$) spectrum of WD\,1647+375 obtained by \citet{gianninasetal11-1} is available in the Montreal White Dwarf Data Base \citep{dufouretal17-1}, which does not exhibit any metal lines. A more recent spectrum ($R\simeq2500$) was obtained as part of the Data Release~1 of the Dark Energy Spectroscopic Instrument \citep[][]{DESI2025}, again not showing any metal lines. 

We therefore observed WD\,1647+375 on 2025 April 23 with the X-shooter spectrograph \citep{vernetetal11-1} mounted on the VLT, using slit widths of 1.0, 0.9, and 0.9\,arcsec and exposure times of 2$\times$200\,s, 2$\times$150\,s and 4$\times$100\,s in the UVB, VIS and NIR arms, respectively. The resulting spectral resolutions are 5400 (UVB), 8900 (VIS), and 5600 (NIR). The data were reduced using the \textsc{reflex} pipeline and the telluric correction was carried out using \textsc{molecfit} \citep{kausch2015}. The only metal absorption line detected in the X-Shooter spectrum is the \ion{Mg}{ii} triplet at 4481\,\AA.

\begin{figure*}
    \centering
    \includegraphics[width=\textwidth]{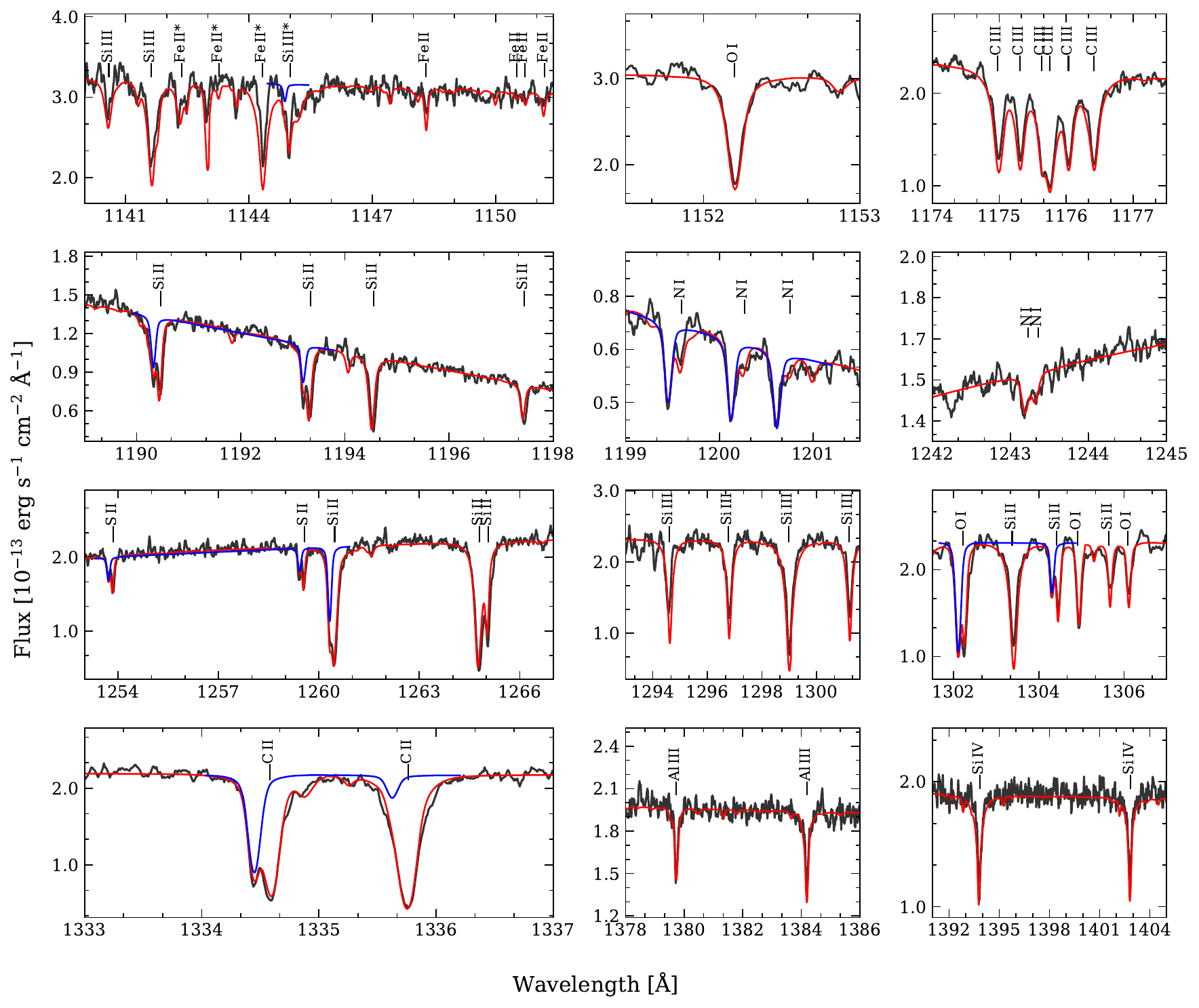}
    \caption{The airglow-corrected deep \textit{HST} COS spectrum of WD\,1647+375 (black) and the best-fit model (red) with the line velocities from Table\,\ref{tab:vel_comp}, and the weighted average abundances from Table\,\ref{tab:abund_params}. Since we have plotted the average abundances, the individual fits to the Si lines look poor. The ISM lines are fitted with a Gaussian profile and are shown in blue. Several transitions of C, N, O, S, Si, Al, and Fe are labelled, where * denotes a blend of Si with Fe lines.}
    \label{fig:hst_spec_fit}
\end{figure*}

\begin{table}
\caption{Summary of parameters for WD\,1647+375. }
\begin{center}
\begin{tabular}{ll}
\hline
\noalign{\smallskip}
Parameters & Value \\
\noalign{\smallskip}
\hline
\noalign{\smallskip}
Name & WD\,J164920.30+372821.25\\
$G$ (mag) & 14.986 (0.003) \\
Parallax (mas) & 12.74 (0.03)\\
\Teff\,(K) & 22\,040 (31) \\
log ($g[\rm cm\,s^{-2}]$) & 7.88 (0.03) \\
$M$ (\Msun)  & 0.57 (0.01)\\
$R$ (\Rsun)  & 0.0144 (0.0003) \\
Cooling age (Myr) & 30 (2) \\
\hline\addlinespace[0.5ex]
\end{tabular}
\end{center}
\footnotesize\small
Note: Astrometric parameters are from \citet{gaiaetal23-1} and spectroscopic parameters are from \citet{sahu2023}.

\label{tab:param}
\end{table}

\begin{table*}
\caption{Summary of abundances of WD\,1647+375. The abundances determined from the two COS spectra are reported separately.}

\begin{center}
\begin{tabular}{lccrrr}
\hline
\noalign{\smallskip}
Element	&	\multicolumn{2}{c}{log\,(Z/H)}	&	$\dot M_\mathrm{Z}$ (\gs)	&	\multicolumn{2}{c}{Mass fraction (\%)}  \\
& snapshot & deep & & WD\,1647+375 & Halley\\
\noalign{\smallskip}
\hline\multicolumn{5}{c}{\textit{HST} COS}\\\hline
\noalign{\smallskip}
C	&	$-$5.22	(0.15)  & $-$5.18	(0.21)	& $1.71\times 10^{7}$	&	9.8 (4.6)	& 24.1 \\
N	&	$-$5.66	(0.29)	& $-$5.68	(0.12)  & $9.23\times10^{6}$	&	5.1 (1.6)	& 1.5 \\
O	&	$-$4.83	(0.04)	& $-$4.69	(0.10)  & $1.28\times10^{8}$	&	67.1 (6.6)	& 35.2 \\
Al	&	$-$6.57	(0.10)	& $-$6.56	(0.10)  &	$1.35\times10^{6}$	&	0.7 (0.2)	& 0.5\\
Si	&	$-$5.76	(0.29)	& $-$5.74	(0.21)  &	$1.01\times10^{7}$	&	5.9 (2.9)	& 12.8 \\
P	&	$<-7.59	$ 		& $<-$7.80  &	$<1.17\times10^{5}$	&		& \\
S	&	$-$6.32	(0.12)	& $-$6.30	(0.10)  &	$5.01\times10^{6}$	&	2.7 (0.8)	& 5.7\\
Fe	&	$-6.20$	(0.15)	& $-$6.20	(0.13)  &	$1.28\times10^{7}$	&	7.0 (2.3)	& 7.2\\
Ni	&	$<-7.11$	 	& $<-$7.56 	&	$<6.45\times10^{5}$	&		& 0.6\\
\noalign{\smallskip}
\hline
\multicolumn{5}{c}{X-shooter}\\\hline
\noalign{\smallskip}
Mg  &      \multicolumn{2}{c}{$-6.12$ (0.14)}    &  $2.99\times10^{6}$ &  1.6 (0.5)   & 6.0 \\
Ca  &      \multicolumn{2}{c}{$<-7.13$}       &  $<7.78\times10^{5}$  &   & 0.6 \\\hline
\noalign{\smallskip}
Total & & & $1.88\times10^{8}$ &\\\hline
\noalign{\smallskip}
\end{tabular}
\end{center}
\footnotesize\small\flushleft
Note: The error in magnesium abundance is adopted from \cite{jamie2024}. The accretion rates and mass fraction are computed using the deep exposure spectra (exposure 2). The mass fraction of heavy elements in Halley's comet \citep{jessberger1988} is provided for comparison. 
\label{tab:abund_params}
\end{table*}

\subsection{Metal abundances}
WD\,1647+375 has $\Teff=22\,040$\,K, $\logg=7.88$, and a hydrogen-dominated atmosphere (Table\,\ref{tab:param}; \citealt{sahu2023}). Fixing these parameters, we generated 1D synthetic spectral model grids for the elements C, N, O, Mg, Al, Si, P, S, Ca, Fe, and Ni using the atmosphere code of \cite{koester2010} but with numerous updates and improvements in the equations of state and absorption coefficients \citep{koester2020}. The model grids span abundances from $-$10 to $-$4\,dex in steps of 0.25\,dex. The analysis of the COS and X-Shooter spectroscopy of WD\,1647+375 followed the methodology described in \cite{Williams2025}.

Since WD\,1647+375 is located at a distance of 78.5\,pc \citep{gaiaetal23-1}, some of the photospheric lines in the COS spectrum are contaminated by absorption lines arising from the ISM, affecting the abundance measurements. We therefore included ISM lines in the fits to the COS spectrum, where each ISM line was modelled with a Gaussian profile using velocity, amplitude, and width as free parameters. The model ISM lines were then combined with the atmospheric model, convolved with the COS line spread function\footnote{\url{https://github.com/spacetelescope/notebooks/blob/master/notebooks/COS/LSF/LSF.ipynb}}, and fitted to the observed spectra using $\chi^2$ minimisation. In these fits, we used a spectral window of 3\,\AA\ centred around the vacuum (COS) and air (X-Shooter) wavelengths of the detected metal lines, and allowed for a local flux scaling factor to match the continua of the data and the model. 

In addition to the ISM lines, the spectral region $1290-1310$\,\AA\ contains many strong photospheric \ion{Si}{ii,iii} lines that affect the level of the continuum and thereby the oxygen abundance determination using \ion{O}{i} 1302.17, 1304.86 and 1306.03\,\AA. To resolve this, we fixed the Si abundance to the initial best-fit value and generated models varying each element relative to Si (one at a time) to find the best-fit measurement. The difference between the abundance determined from the \ion{O}{i} triplet and that from the unblended line at 1152\,\AA\ is comparable to the 1$\sigma$ uncertainties noted in the case of other elements, suggesting that the measured oxygen abundance is reliable.

For nitrogen, we considered the spectral regions covering the \ion{N}{i} triplet at 1200\,\AA\ and the doublet at 1243\,\AA\ which is a blend of four lines with the strongest transitions at 1243.18 and 1243.31\,\AA. We do not detect phosphorus or nickel absorption lines, hence, we determined upper limits considering the transitions \Line{P}{iii}{1344.33}, \Line{Ni}{ii}{1370.12} and following the procedure of \cite{hollands2020}. Figure\,\ref{fig:hst_spec_fit} shows the best-fit model to the spectrum of WD\,1647+375, where the determined abundances are given in Table\,\ref{tab:abund_params}. We note that the ISM lines are resolved from the photospheric lines with average line velocities of $-$19.5 and $+$11\,\kms\ respectively. We computed the final abundance values as the weighted average of the measurements derived from multiple lines of each element, and the uncertainties are adopted as the standard deviation for the COS spectrum.

We determined the best-fit abundances for both COS spectra to (1) probe for temporal variations in the abundances, and (2) assess potential systematic uncertainties. The results are provided in Table\,\ref{tab:abund_params}. We find that the abundances agree well within the $1\sigma$ uncertainties for all elements. There is a difference of 0.14\,dex in the oxygen abundance between the two COS spectra which is most likely related to the different levels of correction for airglow. Overall, the deeper exposure provides the best S/N and thus the smallest uncertainties in detected lines (especially nitrogen, which has only weak lines) and tighter upper limits for non-detections. Hence, the abundance measurements of the deep spectrum were used to study the bulk composition of the accreted body. The results of the snapshot spectrum are provided in the Appendix\,\ref{appendix}.

The snapshot exposure of WD\,1647+375 has been previously analysed \citep{koester2014}, yielding stellar parameters $\Teff=22803\pm310$\,K, $\logg=7.90\pm0.09$, $\log(\mathrm{Si/H})=-6.20\pm0.10$ and $\log(\mathrm{C/H})=-5.50\pm0.20$. We find Si and C abundances larger by 0.4 and 0.3\,dex, respectively, which agree within $2\sigma$ with the earlier work. Using the deep exposure, \citet{wilson2016} derived $\log(\mathrm{C/O})=-1.12\pm0.25$, which is lower than our value of $-0.50\pm0.23$. The abundance differences with these studies are attributed to the newer models, methods, and atomic data used in the current analysis.

\begin{figure}
    \centering
    \includegraphics[width=\columnwidth]{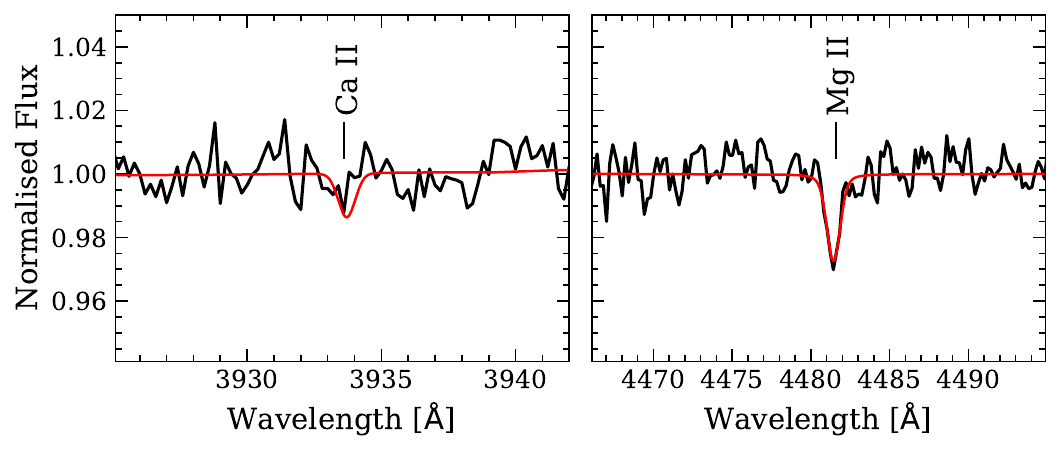}
    \caption{X-shooter spectrum of WD\,1647+375 (black) and the best-fit model (red) with an upper limit of $\log(\rm{Ca/H})$ (left panel) and  $\log(\mathrm{Mg/H})$ (right panel) abundance from Table\,\ref{tab:abund_params}.}
    \label{fig:xshoot_spec_fit}
\end{figure}

The COS spectrum does not contain detectable transitions of magnesium and calcium, and we analysed the X-shooter spectrum to determine the abundances of these two elements. We fitted the spectral region covering the unresolved \ion{Mg}{ii}\,4481\,\AA\ triplet, and \ion{Ca}{ii}\,K. We normalised the observed and model fluxes and varied the model abundances to find a best-fit to the spectrum using $\chi^2$ minimisation. The X-Shooter spectrum and the best-fit model for the \ion{Mg}{ii} are shown in Fig.\,\ref{fig:xshoot_spec_fit}, where we find a systemic velocity of $15.5\pm5.0$\,\kms, consistent with the line velocities from COS. We do not detect \ion{Ca}{ii} absorption and hence we determined an upper limit as outlined above (Table\,\ref{tab:abund_params}).

\subsection{Accretion rates}
The bulk composition of the accreted material can be determined from the measured photospheric abundances by accounting for the different diffusion velocities of the individual elements. As the accreted material diffuses out of the warm and hydrogen-rich atmosphere of WD\,1647+375 on time scales of days \citep{koester2009}, it is safe to assume that there is a steady state between accretion and diffusion, and that the diffusion flux is constant throughout the atmosphere and is equal to the accretion rate, which in turn reflects the elemental composition of the parent body. 

We computed the diffusion fluxes for each detected element following the procedure of \cite{koester2014} which is applicable to warm, hydrogen-atmosphere white dwarfs with $\Teff\geq17\,000\,$K.  Note that the assumption of constant diffusion flux can lead to small changes in the photospheric abundances of the best-fit model \citep{koester2014}. We compared the spectrum emerging from an atmosphere with a constant abundance and a constant diffusion flux, and found no noticeable difference. In addition, we assessed whether radiative levitation may affect the measured abundances. For the \Teff\ and \logg\ of WD\,1643+375, the study by \citet{koester2014} (fig.\,5 and 6) suggests that the effect of radiative levitation is small. We verified this by computing an atmosphere model including radiative levitation for silicon and carbon following the procedures in \cite{koester2014}, and found that the implied changes in the diffusion fluxes are well below the 1$\sigma$ uncertainties in the measurements. The accretion rates for each element are provided in Table\,\ref{tab:abund_params}.

\section{Results and Discussion}
Whereas wind-accretion from a close stellar or substellar companion could explain the volatile-rich nature of the material accreted by WD\,1647+375 \citep{debesetal06-2}, we rule out that scenario for two reasons: (1) \textit{Spitzer} photometry of WD\,1647+375 does not reveal any infrared excess \citep{wilson2019}. (2) Nitrogen is overabundant compared to solar abundances. For example, \cite{gansicke2012} reported COS spectroscopy of three white dwarfs accreting stellar wind from M-dwarf companions and found that the abundances of carbon, oxygen, and sulphur are quasi-solar, with nitrogen remaining below the detection threshold. Mass-loss from sub-stellar companions are expected to be extremely low, with only one optical detection \citep{Walters2023}. This leaves us with the accretion of planetary material as the only option. Below we discuss the composition of the planetary body, as well as its potential nature and origin.  

\subsection{The composition of the planetary body}
\begin{figure}
    \centering
    \includegraphics[width=\columnwidth]{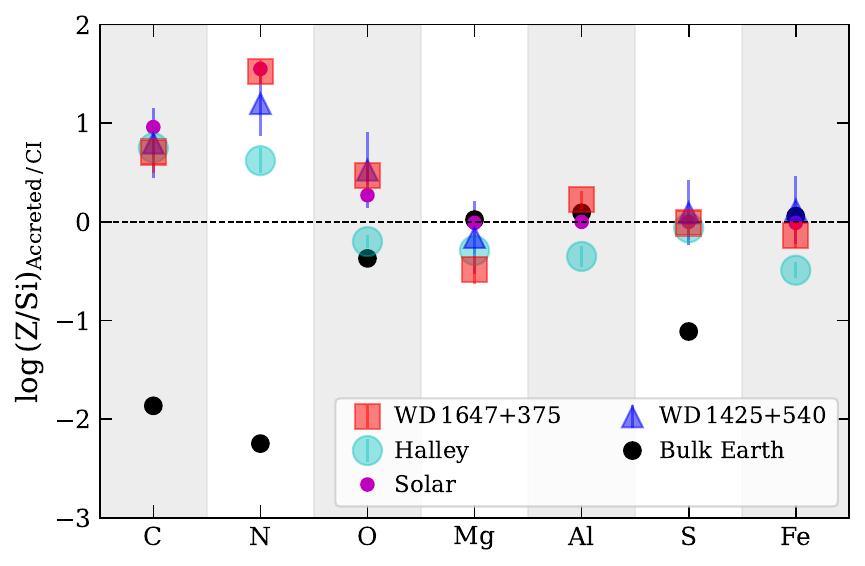}
    \caption{Comparison of log number abundance ratios of WD\,1647+375, relative to silicon and normalised to CI chondrites, with the bulk Earth, Halley's comet \citep{jessberger1988} and WD\,1425+540  (Model 1 from \citealt{xu2017}).}
    \label{fig:che_comp_compare}
\end{figure}

\begin{figure}
    \centering
    \includegraphics[width=1\columnwidth]{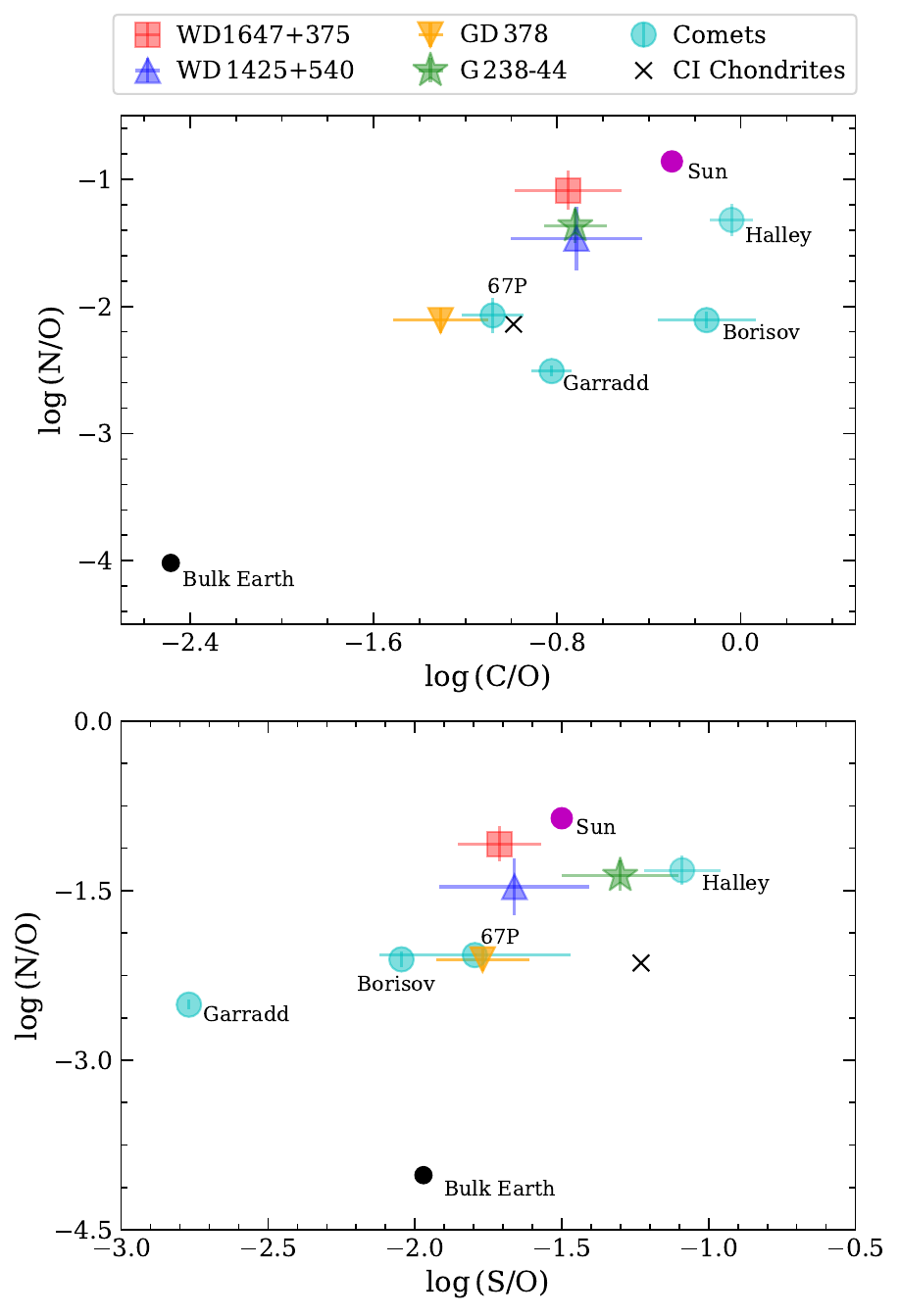}
    \caption{log number abundance ratios of the volatiles carbon, nitrogen, and sulphur relative to oxygen in WD\,1647+375 (red square) compared with the other three white dwarfs that have nitrogen detections (WD\,1425+540\,=\,blue triangle, \citealt{xu2017};  GD\,378\,=\,yellow triangle, \citealt{klein2021}; G\,238-44\,=\,green star, \citealt{johnson2022}), and with well-studied comets (cyan): Halley; 67P/Churyumov-Gerasimenko \citep{rubin2019}; Garradd \citep{paganini2012}; 2I/Borisov \citep{bodewits2020}, CI chondrites, Sun \citep{lodders2003}, and, bulk Earth \citep{MS1995}. Note that 2I/Borisov is an interstellar comet.}
    \label{fig:volatile_ratios}
\end{figure}

We compared the parent body number abundances (see eqn. 2 of \citealt{gansicke2012}) derived from spectroscopy, relative to silicon and normalised to CI chondrites with the abundances of the bulk Earth \citep{MS1995}, comet Halley \citep{jessberger1988}; the Sun \citep{lodders2003}, and WD\,1425+540 (\citet{xu2017}; the first case of volatile-dominated material detected in a polluted white dwarf) as shown in Fig.\,\ref{fig:che_comp_compare}. This comparison suggests that the parent body accreted by WD\,1647+375 closely resembles solar composition, Halley's comet, and the parent body polluting WD\,1425+540. Notably, the accreted planetesimal is rich in carbon, oxygen, and nitrogen, with O/Si and N/Si being $\sim$5 and 8 times higher, respectively than the comet Halley. We further compared the volatile element ratios N/O, C/O, and S/O in Fig.\,\ref{fig:volatile_ratios}, alongside solar system and interstellar comets \citep{paganini2012, rubin2019, bodewits2020}, and the three white dwarfs with photospheric nitrogen detections: WD\,1425+540, G238-44, and GD\,378. Figure\,\ref{fig:volatile_ratios} places WD\,1647+375 closer to the Sun in N/O than to the other, volatile-rich white dwarfs. WD\,1647+375 exhibits $\mathrm{N/O}=0.08$, which is higher than in comets, which have N/O between 0.003 and 0.05. In contrast, the $\mathrm{S/O}=0.02$ of WD\,1647+375 falls within the range observed for the comets 67P and Halley, $\mathrm{S/O} =0.016-0.08$. Overall, the volatile element ratios and high N/O suggest that WD\,1647+375 is accreting icy material analogous to KBOs.

The three white dwarfs with published nitrogen detections span a range of compositions and interpretations: WD\,1425+540 is thought to accrete a KBO analogue \citep{xu2017}, G238-44 two distinct planetary bodies, one being an icy KBO analogue and the other being a rocky body \citep{johnson2022}, and GD378 an icy exo-moon formed in the giant exoplanets' radiation belts (\citealt{Doyle2021}, though recent studies suggest that this hypothesis is unlikely, \citealt{trier2022, Kaiser2025}). 

\begin{figure}
    \centering   
    \includegraphics[width=\columnwidth]{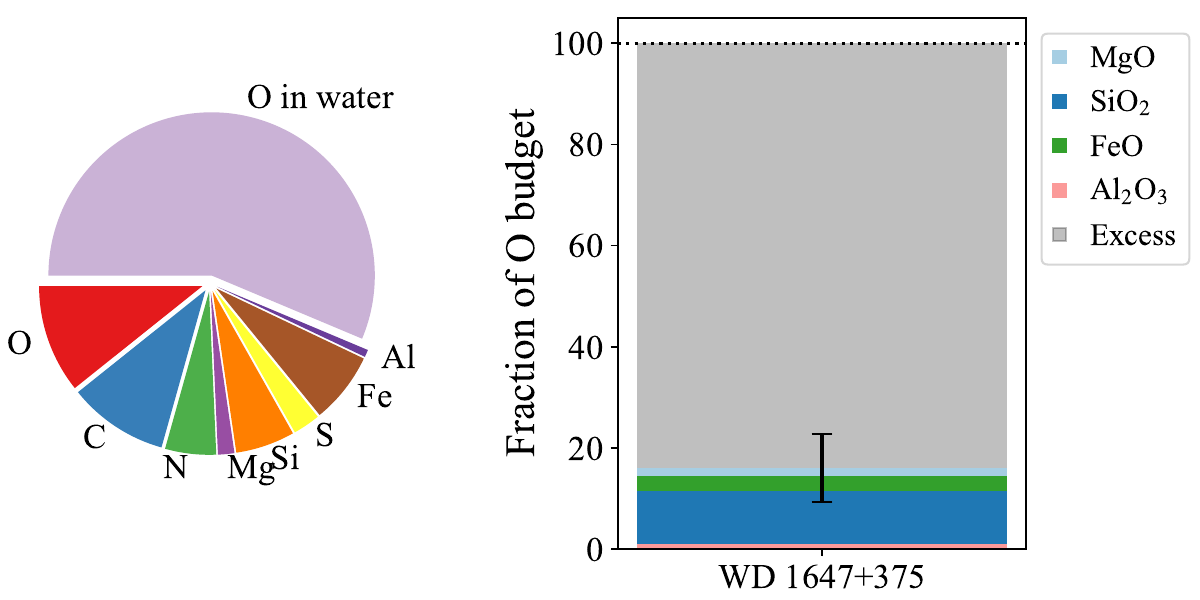}
    \caption{Mass fraction (left) and oxygen budget (right) for the detected heavy elements in WD\,1647+375. The error bar represents the total uncertainty on the oxygen budget.}
    \label{fig:mass_fraction}
\end{figure}

To investigate in more detail the composition of the parent body accreting onto WD\,1647+375, we calculated the mass fractions of the detected elements, which are provided in Table \ref{tab:abund_params} and illustrated in Fig.\,\ref{fig:mass_fraction}. Oxygen is the most abundant element, comprising nearly two-thirds of the total mass. Notably, the volatile nitrogen accounts for $\simeq5$\,per cent of the total mass, the highest value among all white dwarfs with nitrogen detections \citep{xu2017, klein2021, johnson2022}. This nitrogen abundance is also higher than the values determined for the surface of KBOs and other outer solar system bodies \citep{jessberger1988, rubin2019}.

\begin{table}
\caption{Oxygen budget of WD\,1647+375. }
\begin{center}
\begin{tabular}{ll}
\hline
\noalign{\smallskip}
Component & Fraction of oxygen (\%)\\
\hline
Al$_{2}$O$_{3}$ & 1.0 (0.3)\\
SiO$_2$ & 10.4 (6.0)\\
FeO & 3.1 (1.2)\\
MgO & 1.6 (0.5)\\
Excess & 83.9 (7) \\
\hline
\vspace{0.05cm}
\end{tabular}
\end{center}
\label{tab:o_budget}
\end{table}

Following the prescription of \cite{klein2010}, we calculated the water fraction by evaluating the oxygen budget using the standard stoichiometric ratios within common minerals containing the rock-forming elements Mg, Al, Si, and Fe (Table\,\ref{tab:o_budget}). We considered a scenario similar to that of the bulk Earth---namely that oxygen is primarily bound in MgO, Al$_{2}$O$_{3}$, SiO$_{2}$, and FeO. Since Fe can also be metallic (such as in the Earth's core), we allowed the oxidation state of Fe to vary from being all in the form of FeO to all as metallic. Assuming that all Fe is metallic, we find that excess oxygen is $87\pm6\,$per cent. If Fe is entirely locked within FeO, which is a more likely state for a cometary body, the excess oxygen is $84\pm7\,$per cent, as shown in the right panel of Fig.\,\ref{fig:mass_fraction}. This large oxygen excess indicates that the bulk of oxygen was carried as water, and thus the planetary body is rich in water ice. This again supports that WD\,1647+375 is accreting an icy exo-planetesimal.

The water-to-rock ratio is a crucial parameter to constrain the interior structure, composition, and formation history of ice-rich planetary bodies \citep{vazan2022}. From Table\,\ref{tab:wr_frac}, the mass fraction of Mg, Al, Si, and Fe, representative of the rocky material in WD\,1647+375, constitutes $\approx$15\,per cent. From the oxygen budget calculated above, 83.9\,per cent are associated with the accretion of water, and 16.1\,per cent were accreted in the form of minerals. Consequently, the mass fraction of oxygen contained in either water or minerals is 56.4\,per cent and 10.7\,per cent, respectively. The mass fraction of rock is hence the sum of the oxygen and Mg, Al, Si and Fe mass fractions, i.e. 25.9\,per cent. Accounting for the mass fraction of hydrogen associated with the accreted water, 7.1\,per cent, results in a water mass fraction of 63.5\,per cent, and finally in a water-to-rock ratio $2.45\pm0.43$. This ratio is around six times higher compared to the KBOs in our solar system that appear to have water-to-rock ratios of 0.43 \citep{bn2019}.

\begin{table}
\caption{Water and rock mass fraction of icy body accreted by WD\,1647+375.}
\begin{center}
\begin{tabular}{ll}
\hline
Component & mass fraction (\%)\\
\hline
metals (Mg, Al, Si, Fe) & 15.2 (3.7)\\
O in oxides & 10.7 (1.1)\\
O in water (H$_2$O) & 56.4 (5.5)\\
H in water$^*$     & 7.1 (0.7)  \\
\hline
Total H$_2$O & 63.5 (5.6)\\
Total rock (metal+oxide) & 25.9 (3.9)\\
\hline\addlinespace[0.5ex]
\end{tabular}
\end{center}
\footnotesize\small\flushleft
Note: $^*$As the atmosphere of WD\,1647+375 is composed of hydrogen, our observations are not sensitive to the accretion of this element, and therefore it is not included in the mass fractions of the \textit{detected} elements in Table\,\ref{tab:abund_params}. The overall mass fractions including C, N and S (from Table\,\ref{tab:abund_params}) with the above estimates (excluding H in water) amounts to 100 percent.
\label{tab:wr_frac}
\end{table}

\subsection{The nature of the planetary body}
The bulk composition and mass distribution of elements detected in the atmosphere of WD\,1647+375 demonstrate that the accreted object is either a comet or a fragment of the surface of an icy dwarf planet similar to Pluto \citep{Biver2018}. The observed nitrogen enhancement, relative to typical comets, supports the latter scenario, as it can be explained if the accreted material constitutes the mantle and crust of a dwarf planet in a Kuiper-belt location. WD\,1647+375 is accreting at a rate $\approx2\times10^{8}$\,\gs\ for at least 13 years (spanned by our observations). Assuming a water-like density of $\rho=1\,\mathrm{g\,cm^{-3}}$ gives firm lower limits on the mass ($7.7\times10^{16}$\,g) and radius ($2.6$\,km) of the parent body accreting onto the star. If we assume that the accretion phase lasts for $10^5$ to $10^6$\,yr, as estimated from statistical studies of white dwarfs with infrared dust disks \citep{girven2012, cunningham2021, cunningham2025}\footnote{Note that these estimates of the accretion phase duration were derived from the statistical analysis of white dwarfs with dusty circumstellar discs detected in the infrared. It is not evident if the accretion of volatile-dominated material would proceed on the same or similar time scales.}, the parent body would be much larger with a mass of $\sim6\times10^{20-21}$\,g and a radius of about $\simeq50-100$\,km. These values are large compared to solar system comets (e.g., Hale-Bopp's mass is $\sim 10^{19}\,{\rm g}$), but would be typical for a KBO. On the other hand, a shorter accretion phase of $\sim 10^{4}$\,yr would permit a body more comet-like in its mass. 

In the context of the origin of this planetary object, it is interesting to note that the discussions of white dwarfs accreting KBO-like objects \citep{xu2017, johnson2022} implicitly assumed that these bodies were formed in a planetary system around the white dwarf progenitor. However, given that three interstellar comets (1I/Oumuamua, 2I/Borisov, and 3I/ATLAS) were identified in the solar system within a few years \citep{meech2017, jewitt2019, bolin2025}, a tantalizing possibility is that some white dwarfs may accrete such interstellar comets, rather than their own. The likelihood of accreting an interstellar object can be investigated from both dynamical and chemical perspectives.

Dynamically, consider an interstellar comet, composed of many particles, that is initially in a hyperbolic orbit around a white dwarf. This orbit may be changed by non-gravitational decelerations on approach to periastron, assuming that the interstellar object contains volatiles to be released \citep{VEG2015} as noted in the cometary orbits in the solar system. Independent of these details, if the comet passes through the white dwarf Roche sphere, which has an approximate radius of $R_{\rm Roche}=1\mathrm{R}_{\odot}$, it will tidally disrupt. This process is similar to the well-studied scenario of stars being tidally disrupted by black holes, where roughly half of the material remains bound, and the other half is ejected \citep{lacy1982, Rees1988}. Numerical simulations have demonstrated that a similar ratio of bound to ejected material is true for the tidal disruption of solid planetary bodies \citep{veras2014, MP2020}.The resulting tidal disruption of the interstellar object would create a debris disc that eventually is accreted onto the white dwarf over a timescale $t_{\rm disc}$. The upper bound of $t_{\rm disc}$ is, as stated above, $\sim$ Myr \citep{girven2012, cunningham2021, cunningham2025}, while the lower bound is poorly constrained \citep{verhen2020}. Then the expected number of accretion events per white dwarf is $\sim (\pi R_{\rm Roche}^2) n v t_{\rm disc}$, where $v$ is the stellar velocity dispersion in the relevant local patch of the Milky Way, and $n$ is the number density of interstellar objects. The values of $n$ and $t_{\rm disc}$ are highly uncertain. However, generating an interstellar object pollution rate of one out of every few hundred white dwarfs can be obtained with reasonable assumptions (e.g. $n=0.2\,\mathrm{au}^{-3}$; \citealt{doetal2018}, $t_{\rm disc} = 10^2$~yr, and $v=10\,\mathrm{km\,s}^{-1}$). For comparison, \cite{dehhansch2022} estimate the rate of interstellar bodies hitting the Sun as 17 in 1000\,yr.

Alternatively, an interstellar object may be gravitationally captured by the white dwarf \citep{napadabat2021a,napadabat2021b,dehhan2022,dehhansch2022,belgre2024}. Once in the system, the object could experience a combination of non-gravitational forces and close encounters with any existing planets, eventually leading to accretion onto the white dwarf \citep{bonwya2012,wyaetal2017,maretal2018,rodlai2024}. However, the capture rate of interstellar objects by a Sun-Jupiter like system is small ($\sim2$ objects per 1000\,yr) and can only occur at low incoming speeds (typically $v_{\infty} \leq 4\,\mathrm{km\,s}^{-1}$ \citealt{dehhansch2022}). 

Chemically, 2I/Borisov has many characteristics of a KBO \citep{jewitt2023}. The nature of 1I/'Oumuamua is less well understood, but is either likely to be a small (diameter $< 100$\,m) fragment of N$_2$ ice from the surface of an exo-Pluto \citep{desch2021} or a dark comet arising from the cold and distant regions of the extra-solar system \citep{seligman2022}. There are similar fragments of crusts of differentiated Pluto-like dwarf planets observed in our solar system \citep{Biver2018}. For example, the comet C/2016 R2 is thought to sample the N$_2$ ice-rich crust of a differentiated Pluto-like body, and such larger objects could be expected among comets ejected from stellar systems. Thus, compositionally, interstellar objects are likely to resemble other intrinsic comets, but there is a distinct possibility, based on observations like C/2016R2 in our solar system, that some interstellar comets could be N$_2$ rich. The chemical composition of the recently discovered interstellar comet, 3I/ATLAS is not yet sufficiently well-constrained to be discussed within the context of our study of WD\,1647+375  \citep{puzia2025,jewitt2025}.
Nevertheless, regardless of their actual origin, white dwarfs have the potential to provide an important tool for measuring the abundances of icy exo-planetesimals. 

\section{Conclusion}
The detection of nitrogen alongside other heavy elements indicates that WD\,1647+375 is accreting an icy, Kuiper Belt-like body with a large (63.5\,per cent) water content. WD\,1647+375 is the first hydrogen atmosphere white dwarf confirmed to be purely accreting a KBO analogue, similar to the helium atmosphere white dwarf WD\,1425+540. Given the fact that helium-rich atmospheres have sinking timescales orders of magnitude longer than their hydrogen-rich counterparts, and since their accretion history is therefore often uncertain \citep{koester2009, obrienetal25-1}, the discovery of WD\,1647+375 provides an unambiguous measurement of metal abundances and accretion rates for a planetary body with comet-like composition.

The origin of the body accreting onto WD\,1647+375 remains uncertain, and we raise the possibility that it may be of interstellar origin, rather than belonging to a planetary system formed around the progenitor of the white dwarf. The relative probabilities of different scenarios (an intrinsic comet, a KBO fragment, or an interstellar comet) are subject to a number of uncertain parameters, including the fraction of intrinsic comets lost during the (post) main-sequence evolution, and the space density, mass distribution, and lifetime of interstellar comets. It is difficult to distinguish between these scenarios solely from its chemical composition, and it will require more detailed dynamical simulations to explore the likelihood of the different possible origins.

Our study of WD\,1647+375 demonstrates the existence of icy exo-planetesimals that could deliver water and other volatiles to terrestrial planets in extrasolar systems~--~a pre-requisite for the development of life in other worlds. This discovery is based on the far-ultraviolet spectroscopic capabilities of \textit{HST}, emphasising the importance for future ultraviolet missions in exploring other worlds and the origins of life.

\bibliographystyle{mnras}
\bibliography{ref, aabib, proceedings+books}

\begin{thebibliography}{}
\makeatletter
\relax
\def\mn@urlcharsother{\let\do\@makeother \do\$\do\&\do\#\do\^\do\_\do\%\do\~}
\def\mn@doi{\begingroup\mn@urlcharsother \@ifnextchar [ {\mn@doi@} {\mn@doi@[]}}
\def\mn@doi@[#1]#2{\def\@tempa{#1}\ifx\@tempa\@empty \href {http://dx.doi.org/#2} {doi:#2}\else \href {http://dx.doi.org/#2} {#1}\fi \endgroup}
\def\mn@eprint#1#2{\mn@eprint@#1:#2::\@nil}
\def\mn@eprint@arXiv#1{\href {http://arxiv.org/abs/#1} {{\tt arXiv:#1}}}
\def\mn@eprint@dblp#1{\href {http://dblp.uni-trier.de/rec/bibtex/#1.xml} {dblp:#1}}
\def\mn@eprint@#1:#2:#3:#4\@nil{\def\@tempa {#1}\def\@tempb {#2}\def\@tempc {#3}\ifx \@tempc \@empty \let \@tempc \@tempb \let \@tempb \@tempa \fi \ifx \@tempb \@empty \def\@tempb {arXiv}\fi \@ifundefined {mn@eprint@\@tempb}{\@tempb:\@tempc}{\expandafter \expandafter \csname mn@eprint@\@tempb\endcsname \expandafter{\@tempc}}}

\bibitem[\protect\citeauthoryear{{Anders}}{{Anders}}{1989}]{anders89-1}
{Anders} E.,  1989, \mn@doi [Nature] {10.1038/342255a0}, \href {https://ui.adsabs.harvard.edu/abs/1989Natur.342..255A} {342, 255}

\bibitem[\protect\citeauthoryear{{Anslow}, {Bonsor}  \& {Rimmer}}{{Anslow} et~al.}{2023}]{anslowetal23-1}
{Anslow} R.~J.,  {Bonsor} A.,   {Rimmer} P.~B.,  2023, \mn@doi [Proceedings of the Royal Society of London Series A] {10.1098/rspa.2023.0434}, \href {https://ui.adsabs.harvard.edu/abs/2023RSPSA.47930434A} {479, 20230434}

\bibitem[\protect\citeauthoryear{{Barrett}, {Bryson}  \& {Geraki}}{{Barrett} et~al.}{2025}]{barrettetal25-1}
{Barrett} T.~J.,  {Bryson} J. F.~J.,   {Geraki} K.,  2025, \mn@doi [\icarus] {10.1016/j.icarus.2025.116588}, \href {https://ui.adsabs.harvard.edu/abs/2025Icar..43616588B} {436, 116588}

\bibitem[\protect\citeauthoryear{{Belbruno} \& {Green}}{{Belbruno} \& {Green}}{2024}]{belgre2024}
{Belbruno} E.,  {Green} J.,  2024, \mn@doi [Celestial Mechanics and Dynamical Astronomy] {10.1007/s10569-024-10223-1}, \href {https://ui.adsabs.harvard.edu/abs/2024CeMDA.136...52B} {136, 52}

\bibitem[\protect\citeauthoryear{{Bierson} \& {Nimmo}}{{Bierson} \& {Nimmo}}{2019}]{bn2019}
{Bierson} C.~J.,  {Nimmo} F.,  2019, \mn@doi [\icarus] {10.1016/j.icarus.2019.01.027}, \href {https://ui.adsabs.harvard.edu/abs/2019Icar..326...10B} {326, 10}

\bibitem[\protect\citeauthoryear{{Biver} et~al.,}{{Biver} et~al.}{2018}]{Biver2018}
{Biver} N.,  et~al., 2018, \mn@doi [\aap] {10.1051/0004-6361/201833449}, \href {https://ui.adsabs.harvard.edu/abs/2018A&A...619A.127B} {619, A127}

\bibitem[\protect\citeauthoryear{{Bodewits} et~al.,}{{Bodewits} et~al.}{2020}]{bodewits2020}
{Bodewits} D.,  et~al., 2020, \mn@doi [Nature Astronomy] {10.1038/s41550-020-1095-2}, \href {https://ui.adsabs.harvard.edu/abs/2020NatAs...4..867B} {4, 867}

\bibitem[\protect\citeauthoryear{{Bolin} et~al.,}{{Bolin} et~al.}{2025}]{bolin2025}
{Bolin} B.~T.,  et~al., 2025, \mn@doi [\mnras] {10.1093/mnrasl/slaf078}, \href {https://ui.adsabs.harvard.edu/abs/2025MNRAS.tmpL..75B} {}

\bibitem[\protect\citeauthoryear{{Bonsor} \& {Wyatt}}{{Bonsor} \& {Wyatt}}{2012}]{bonwya2012}
{Bonsor} A.,  {Wyatt} M.~C.,  2012, \mn@doi [\mnras] {10.1111/j.1365-2966.2011.20156.x}, \href {https://ui.adsabs.harvard.edu/abs/2012MNRAS.420.2990B} {420, 2990}

\bibitem[\protect\citeauthoryear{{Bourrier} et~al.,}{{Bourrier} et~al.}{2018}]{bourier2018}
{Bourrier} V.,  et~al., 2018, \mn@doi [\aap] {10.1051/0004-6361/201832700}, \href {https://ui.adsabs.harvard.edu/abs/2018A&A...615A.117B} {615, A117}

\bibitem[\protect\citeauthoryear{{Brouwers}, {Buchan}, {Bonsor}, {Malamud}, {Lynch}, {Rogers}  \& {Koester}}{{Brouwers} et~al.}{2023}]{brouwers2023}
{Brouwers} M.~G.,  {Buchan} A.~M.,  {Bonsor} A.,  {Malamud} U.,  {Lynch} E.,  {Rogers} L.,   {Koester} D.,  2023, \mn@doi [\mnras] {10.1093/mnras/stac3317}, \href {https://ui.adsabs.harvard.edu/abs/2023MNRAS.519.2663B} {519, 2663}

\bibitem[\protect\citeauthoryear{{Cunningham} et~al.,}{{Cunningham} et~al.}{2021}]{cunningham2021}
{Cunningham} T.,  et~al., 2021, \mn@doi [\mnras] {10.1093/mnras/stab553}, \href {https://ui.adsabs.harvard.edu/abs/2021MNRAS.503.1646C} {503, 1646}

\bibitem[\protect\citeauthoryear{{Cunningham} et~al.,}{{Cunningham} et~al.}{2025}]{cunningham2025}
{Cunningham} T.,  et~al., 2025, \mn@doi [\mnras] {10.1093/mnras/staf428}, \href {https://ui.adsabs.harvard.edu/abs/2025MNRAS.539.2021C} {539, 2021}

\bibitem[\protect\citeauthoryear{{DESI Collaboration} et~al.,}{{DESI Collaboration} et~al.}{2025}]{DESI2025}
{DESI Collaboration} et~al., 2025, \mn@doi [arXiv e-prints] {10.48550/arXiv.2503.14745}, \href {https://ui.adsabs.harvard.edu/abs/2025arXiv250314745D} {p. arXiv:2503.14745}

\bibitem[\protect\citeauthoryear{{Debes}}{{Debes}}{2006}]{debesetal06-2}
{Debes} J.~H.,  2006, \mn@doi [ApJ] {10.1086/508132}, \href {https://ui.adsabs.harvard.edu/abs/2006ApJ...652..636D} {652, 636}

\bibitem[\protect\citeauthoryear{{Dehnen} \& {Hands}}{{Dehnen} \& {Hands}}{2022}]{dehhan2022}
{Dehnen} W.,  {Hands} T.~O.,  2022, \mn@doi [\mnras] {10.1093/mnras/stab3670}, \href {https://ui.adsabs.harvard.edu/abs/2022MNRAS.512.4062D} {512, 4062}

\bibitem[\protect\citeauthoryear{{Dehnen}, {Hands}  \& {Sch{\"o}nrich}}{{Dehnen} et~al.}{2022}]{dehhansch2022}
{Dehnen} W.,  {Hands} T.~O.,   {Sch{\"o}nrich} R.,  2022, \mn@doi [\mnras] {10.1093/mnras/stab3666}, \href {https://ui.adsabs.harvard.edu/abs/2022MNRAS.512.4078D} {512, 4078}

\bibitem[\protect\citeauthoryear{{Desch} \& {Jackson}}{{Desch} \& {Jackson}}{2021}]{desch2021}
{Desch} S.~J.,  {Jackson} A.~P.,  2021, \mn@doi [Journal of Geophysical Research (Planets)] {10.1029/2020JE006807}, \href {https://ui.adsabs.harvard.edu/abs/2021JGRE..12606807D} {126, e06807}

\bibitem[\protect\citeauthoryear{{Do}, {Tucker}  \& {Tonry}}{{Do} et~al.}{2018}]{doetal2018}
{Do} A.,  {Tucker} M.~A.,   {Tonry} J.,  2018, \mn@doi [\apjl] {10.3847/2041-8213/aaae67}, \href {https://ui.adsabs.harvard.edu/abs/2018ApJ...855L..10D} {855, L10}

\bibitem[\protect\citeauthoryear{{Doyle}, {Desch}  \& {Young}}{{Doyle} et~al.}{2021}]{Doyle2021}
{Doyle} A.~E.,  {Desch} S.~J.,   {Young} E.~D.,  2021, \mn@doi [\apjl] {10.3847/2041-8213/abd9ba}, \href {https://ui.adsabs.harvard.edu/abs/2021ApJ...907L..35D} {907, L35}

\bibitem[\protect\citeauthoryear{{Doyle} et~al.,}{{Doyle} et~al.}{2023}]{doyleetal23-1}
{Doyle} A.~E.,  et~al., 2023, \mn@doi [ApJ] {10.3847/1538-4357/acbd44}, \href {https://ui.adsabs.harvard.edu/abs/2023ApJ...950...93D} {950, 93}

\bibitem[\protect\citeauthoryear{{Dufour}, {Blouin}, {Coutu}, {Fortin-Archambault}, {Thibeault}, {Bergeron}  \& {Fontaine}}{{Dufour} et~al.}{2017}]{dufouretal17-1}
{Dufour} P.,  {Blouin} S.,  {Coutu} S.,  {Fortin-Archambault} M.,  {Thibeault} C.,  {Bergeron} P.,   {Fontaine} G.,  2017, in {Tremblay} P.-E.,  {G\"ansicke} B.~T.,   {Marsh} T.~R.,  eds, 20th European Workshop on White Dwarfs. ASP Conf. Ser. 509

\bibitem[\protect\citeauthoryear{{Farihi}, {Barstow}, {Redfield}, {Dufour}  \& {Hambly}}{{Farihi} et~al.}{2010}]{farihi2010}
{Farihi} J.,  {Barstow} M.~A.,  {Redfield} S.,  {Dufour} P.,   {Hambly} N.~C.,  2010, \mn@doi [\mnras] {10.1111/j.1365-2966.2010.16426.x}, \href {https://ui.adsabs.harvard.edu/abs/2010MNRAS.404.2123F} {404, 2123}

\bibitem[\protect\citeauthoryear{{Farihi}, {G{\"a}nsicke}  \& {Koester}}{{Farihi} et~al.}{2013}]{farihi2013}
{Farihi} J.,  {G{\"a}nsicke} B.~T.,   {Koester} D.,  2013, \mn@doi [Science] {10.1126/science.1239447}, \href {https://ui.adsabs.harvard.edu/abs/2013Sci...342..218F} {342, 218}

\bibitem[\protect\citeauthoryear{{Gaia Collaboration} et~al.,}{{Gaia Collaboration} et~al.}{2023}]{gaiaetal23-1}
{Gaia Collaboration} et~al., 2023, \mn@doi [A\&A] {10.1051/0004-6361/202243940}, \href {https://ui.adsabs.harvard.edu/abs/2023A&A...674A...1G} {674, A1}

\bibitem[\protect\citeauthoryear{{G{\"a}nsicke}, {Koester}, {Farihi}, {Girven}, {Parsons}  \& {Breedt}}{{G{\"a}nsicke} et~al.}{2012}]{gansicke2012}
{G{\"a}nsicke} B.~T.,  {Koester} D.,  {Farihi} J.,  {Girven} J.,  {Parsons} S.~G.,   {Breedt} E.,  2012, \mn@doi [\mnras] {10.1111/j.1365-2966.2012.21201.x}, \href {https://ui.adsabs.harvard.edu/abs/2012MNRAS.424..333G} {424, 333}

\bibitem[\protect\citeauthoryear{{Gentile Fusillo}, {G{\"a}nsicke}, {Farihi}, {Koester}, {Schreiber}  \& {Pala}}{{Gentile Fusillo} et~al.}{2017}]{GF2017}
{Gentile Fusillo} N.~P.,  {G{\"a}nsicke} B.~T.,  {Farihi} J.,  {Koester} D.,  {Schreiber} M.~R.,   {Pala} A.~F.,  2017, \mn@doi [\mnras] {10.1093/mnras/stx468}, \href {https://ui.adsabs.harvard.edu/abs/2017MNRAS.468..971G} {468, 971}

\bibitem[\protect\citeauthoryear{{Gianninas}, {Bergeron}  \& {Ruiz}}{{Gianninas} et~al.}{2011}]{gianninasetal11-1}
{Gianninas} A.,  {Bergeron} P.,   {Ruiz} M.~T.,  2011, \mn@doi [ApJ] {10.1088/0004-637X/743/2/138}, \href {https://ui.adsabs.harvard.edu/abs/2011ApJ...743..138G} {743, 138}

\bibitem[\protect\citeauthoryear{{Girven}, {Brinkworth}, {Farihi}, {G{\"a}nsicke}, {Hoard}, {Marsh}  \& {Koester}}{{Girven} et~al.}{2012}]{girven2012}
{Girven} J.,  {Brinkworth} C.~S.,  {Farihi} J.,  {G{\"a}nsicke} B.~T.,  {Hoard} D.~W.,  {Marsh} T.~R.,   {Koester} D.,  2012, \mn@doi [\apj] {10.1088/0004-637X/749/2/154}, \href {https://ui.adsabs.harvard.edu/abs/2012ApJ...749..154G} {749, 154}

\bibitem[\protect\citeauthoryear{{Hollands}, {Koester}, {Alekseev}, {Herbert}  \& {G{\"a}nsicke}}{{Hollands} et~al.}{2017}]{hollands2017}
{Hollands} M.~A.,  {Koester} D.,  {Alekseev} V.,  {Herbert} E.~L.,   {G{\"a}nsicke} B.~T.,  2017, \mn@doi [\mnras] {10.1093/mnras/stx250}, \href {https://ui.adsabs.harvard.edu/abs/2017MNRAS.467.4970H} {467, 4970}

\bibitem[\protect\citeauthoryear{{Hollands} et~al.,}{{Hollands} et~al.}{2020}]{hollands2020}
{Hollands} M.~A.,  et~al., 2020, \mn@doi [Nature Astronomy] {10.1038/s41550-020-1028-0}, \href {https://ui.adsabs.harvard.edu/abs/2020NatAs...4..663H} {4, 663}

\bibitem[\protect\citeauthoryear{{Izquierdo}, {Toloza}, {G{\"a}nsicke}, {Rodr{\'\i}guez-Gil}, {Farihi}, {Koester}, {Guo}  \& {Redfield}}{{Izquierdo} et~al.}{2021}]{pi2021}
{Izquierdo} P.,  {Toloza} O.,  {G{\"a}nsicke} B.~T.,  {Rodr{\'\i}guez-Gil} P.,  {Farihi} J.,  {Koester} D.,  {Guo} J.,   {Redfield} S.,  2021, \mn@doi [\mnras] {10.1093/mnras/staa3987}, \href {https://ui.adsabs.harvard.edu/abs/2021MNRAS.501.4276I} {501, 4276}

\bibitem[\protect\citeauthoryear{{Jackson} \& {Desch}}{{Jackson} \& {Desch}}{2021}]{jackson2021}
{Jackson} A.~P.,  {Desch} S.~J.,  2021, \mn@doi [Journal of Geophysical Research (Planets)] {10.1029/2020JE006706}, \href {https://ui.adsabs.harvard.edu/abs/2021JGRE..12606706J} {126, e06706}

\bibitem[\protect\citeauthoryear{{Jessberger}, {Christoforidis}  \& {Kissel}}{{Jessberger} et~al.}{1988}]{jessberger1988}
{Jessberger} E.~K.,  {Christoforidis} A.,   {Kissel} J.,  1988, \mn@doi [\nat] {10.1038/332691a0}, \href {https://ui.adsabs.harvard.edu/abs/1988Natur.332..691J} {332, 691}

\bibitem[\protect\citeauthoryear{{Jewitt} \& {Luu}}{{Jewitt} \& {Luu}}{2019}]{jewitt2019}
{Jewitt} D.,  {Luu} J.,  2019, \mn@doi [\apjl] {10.3847/2041-8213/ab530b}, \href {https://ui.adsabs.harvard.edu/abs/2019ApJ...886L..29J} {886, L29}

\bibitem[\protect\citeauthoryear{{Jewitt} \& {Seligman}}{{Jewitt} \& {Seligman}}{2023}]{jewitt2023}
{Jewitt} D.,  {Seligman} D.~Z.,  2023, \mn@doi [\araa] {10.1146/annurev-astro-071221-054221}, \href {https://ui.adsabs.harvard.edu/abs/2023ARA&A..61..197J} {61, 197}

\bibitem[\protect\citeauthoryear{{Jewitt}, {Hui}, {Mutchler}, {Kim}  \& {Agarwal}}{{Jewitt} et~al.}{2025}]{jewitt2025}
{Jewitt} D.,  {Hui} M.-T.,  {Mutchler} M.,  {Kim} Y.,   {Agarwal} J.,  2025, \mn@doi [arXiv e-prints] {10.48550/arXiv.2508.02934}, \href {https://ui.adsabs.harvard.edu/abs/2025arXiv250802934J} {p. arXiv:2508.02934}

\bibitem[\protect\citeauthoryear{{Johnson}, {Klein}, {Koester}, {Melis}, {Zuckerman}  \& {Jura}}{{Johnson} et~al.}{2022}]{johnson2022}
{Johnson} T.~M.,  {Klein} B.~L.,  {Koester} D.,  {Melis} C.,  {Zuckerman} B.,   {Jura} M.,  2022, \mn@doi [\apj] {10.3847/1538-4357/aca089}, \href {https://ui.adsabs.harvard.edu/abs/2022ApJ...941..113J} {941, 113}

\bibitem[\protect\citeauthoryear{{Jura}}{{Jura}}{2003}]{Jura2003}
{Jura} M.,  2003, \mn@doi [\apjl] {10.1086/374036}, \href {https://ui.adsabs.harvard.edu/abs/2003ApJ...584L..91J} {584, L91}

\bibitem[\protect\citeauthoryear{{Kaiser}, {Clemens}, {Blouin}, {Dennihy}, {Dufour}, {Hegedus}  \& {Reding}}{{Kaiser} et~al.}{2025}]{Kaiser2025}
{Kaiser} B.~C.,  {Clemens} J.~C.,  {Blouin} S.,  {Dennihy} E.,  {Dufour} P.,  {Hegedus} R.~J.,   {Reding} J.~S.,  2025, \mn@doi [\apj] {10.3847/1538-4357/ad9a6d}, \href {https://ui.adsabs.harvard.edu/abs/2025ApJ...979..111K} {979, 111}

\bibitem[\protect\citeauthoryear{{Kausch} et~al.,}{{Kausch} et~al.}{2015}]{kausch2015}
{Kausch} W.,  et~al., 2015, \mn@doi [\aap] {10.1051/0004-6361/201423909}, \href {https://ui.adsabs.harvard.edu/abs/2015A&A...576A..78K} {576, A78}

\bibitem[\protect\citeauthoryear{{Kawka} \& {Vennes}}{{Kawka} \& {Vennes}}{2012}]{kawka2012}
{Kawka} A.,  {Vennes} S.,  2012, \mn@doi [\aap] {10.1051/0004-6361/201118210}, \href {https://ui.adsabs.harvard.edu/abs/2012A&A...538A..13K} {538, A13}

\bibitem[\protect\citeauthoryear{{Kiefer}, {Lecavelier des Etangs}, {Boissier}, {Vidal-Madjar}, {Beust}, {Lagrange}, {H{\'e}brard}  \& {Ferlet}}{{Kiefer} et~al.}{2014a}]{keifer2014}
{Kiefer} F.,  {Lecavelier des Etangs} A.,  {Boissier} J.,  {Vidal-Madjar} A.,  {Beust} H.,  {Lagrange} A.~M.,  {H{\'e}brard} G.,   {Ferlet} R.,  2014a, \mn@doi [\nat] {10.1038/nature13849}, \href {https://ui.adsabs.harvard.edu/abs/2014Natur.514..462K} {514, 462}

\bibitem[\protect\citeauthoryear{{Kiefer}, {Lecavelier des Etangs}, {Augereau}, {Vidal-Madjar}, {Lagrange}  \& {Beust}}{{Kiefer} et~al.}{2014b}]{keifer2014b}
{Kiefer} F.,  {Lecavelier des Etangs} A.,  {Augereau} J.~C.,  {Vidal-Madjar} A.,  {Lagrange} A.~M.,   {Beust} H.,  2014b, \mn@doi [\aap] {10.1051/0004-6361/201323128}, \href {https://ui.adsabs.harvard.edu/abs/2014A&A...561L..10K} {561, L10}

\bibitem[\protect\citeauthoryear{{Klein}, {Jura}, {Koester}, {Zuckerman}  \& {Melis}}{{Klein} et~al.}{2010}]{klein2010}
{Klein} B.,  {Jura} M.,  {Koester} D.,  {Zuckerman} B.,   {Melis} C.,  2010, \mn@doi [\apj] {10.1088/0004-637X/709/2/950}, \href {https://ui.adsabs.harvard.edu/abs/2010ApJ...709..950K} {709, 950}

\bibitem[\protect\citeauthoryear{{Klein}, {Doyle}, {Zuckerman}, {Dufour}, {Blouin}, {Melis}, {Weinberger}  \& {Young}}{{Klein} et~al.}{2021}]{klein2021}
{Klein} B.~L.,  {Doyle} A.~E.,  {Zuckerman} B.,  {Dufour} P.,  {Blouin} S.,  {Melis} C.,  {Weinberger} A.~J.,   {Young} E.~D.,  2021, \mn@doi [\apj] {10.3847/1538-4357/abe40b}, \href {https://ui.adsabs.harvard.edu/abs/2021ApJ...914...61K} {914, 61}

\bibitem[\protect\citeauthoryear{{Koester}}{{Koester}}{2009}]{koester2009}
{Koester} D.,  2009, \mn@doi [\aap] {10.1051/0004-6361/200811468}, \href {https://ui.adsabs.harvard.edu/abs/2009A&A...498..517K} {498, 517}

\bibitem[\protect\citeauthoryear{{Koester}}{{Koester}}{2010}]{koester2010}
{Koester} D.,  2010, \memsai, \href {https://ui.adsabs.harvard.edu/abs/2010MmSAI..81..921K} {81, 921}

\bibitem[\protect\citeauthoryear{{Koester}, {G{\"a}nsicke}  \& {Farihi}}{{Koester} et~al.}{2014}]{koester2014}
{Koester} D.,  {G{\"a}nsicke} B.~T.,   {Farihi} J.,  2014, \mn@doi [\aap] {10.1051/0004-6361/201423691}, \href {https://ui.adsabs.harvard.edu/abs/2014A&A...566A..34K} {566, A34}

\bibitem[\protect\citeauthoryear{{Koester}, {Kepler}  \& {Irwin}}{{Koester} et~al.}{2020}]{koester2020}
{Koester} D.,  {Kepler} S.~O.,   {Irwin} A.~W.,  2020, \mn@doi [\aap] {10.1051/0004-6361/202037530}, \href {https://ui.adsabs.harvard.edu/abs/2020A&A...635A.103K} {635, A103}

\bibitem[\protect\citeauthoryear{{Kral}, {Huet}, {Bergez-Casalou}, {Th{\'e}bault}, {Charnoz}  \& {Fornasier}}{{Kral} et~al.}{2024}]{kraletal24-1}
{Kral} Q.,  {Huet} P.,  {Bergez-Casalou} C.,  {Th{\'e}bault} P.,  {Charnoz} S.,   {Fornasier} S.,  2024, \mn@doi [A\&A] {10.1051/0004-6361/202451263}, \href {https://ui.adsabs.harvard.edu/abs/2024A&A...692A..70K} {692, A70}

\bibitem[\protect\citeauthoryear{{Lacy}, {Townes}  \& {Hollenbach}}{{Lacy} et~al.}{1982}]{lacy1982}
{Lacy} J.~H.,  {Townes} C.~H.,   {Hollenbach} D.~J.,  1982, \mn@doi [\apj] {10.1086/160402}, \href {https://ui.adsabs.harvard.edu/abs/1982ApJ...262..120L} {262, 120}

\bibitem[\protect\citeauthoryear{{Lodders}}{{Lodders}}{2003}]{lodders2003}
{Lodders} K.,  2003, \mn@doi [\apj] {10.1086/375492}, \href {https://ui.adsabs.harvard.edu/abs/2003ApJ...591.1220L} {591, 1220}

\bibitem[\protect\citeauthoryear{{Malamud} \& {Perets}}{{Malamud} \& {Perets}}{2016}]{malamud2016}
{Malamud} U.,  {Perets} H.~B.,  2016, \mn@doi [\apj] {10.3847/0004-637X/832/2/160}, \href {https://ui.adsabs.harvard.edu/abs/2016ApJ...832..160M} {832, 160}

\bibitem[\protect\citeauthoryear{{Malamud} \& {Perets}}{{Malamud} \& {Perets}}{2020}]{MP2020}
{Malamud} U.,  {Perets} H.~B.,  2020, \mn@doi [\mnras] {10.1093/mnras/staa142}, \href {https://ui.adsabs.harvard.edu/abs/2020MNRAS.492.5561M} {492, 5561}

\bibitem[\protect\citeauthoryear{{Marino}, {Bonsor}, {Wyatt}  \& {Kral}}{{Marino} et~al.}{2018}]{maretal2018}
{Marino} S.,  {Bonsor} A.,  {Wyatt} M.~C.,   {Kral} Q.,  2018, \mn@doi [\mnras] {10.1093/mnras/sty1475}, \href {https://ui.adsabs.harvard.edu/abs/2018MNRAS.479.1651M} {479, 1651}

\bibitem[\protect\citeauthoryear{{McDonough} \& {Sun}}{{McDonough} \& {Sun}}{1995}]{MS1995}
{McDonough} W.~F.,  {Sun} S.~s.,  1995, \mn@doi [Chemical Geology] {10.1016/0009-2541(94)00140-4}, \href {https://ui.adsabs.harvard.edu/abs/1995ChGeo.120..223M} {120, 223}

\bibitem[\protect\citeauthoryear{{Meech} et~al.,}{{Meech} et~al.}{2017}]{meech2017}
{Meech} K.~J.,  et~al., 2017, \mn@doi [\nat] {10.1038/nature25020}, \href {https://ui.adsabs.harvard.edu/abs/2017Natur.552..378M} {552, 378}

\bibitem[\protect\citeauthoryear{{Melis} \& {Dufour}}{{Melis} \& {Dufour}}{2017}]{dufour2017}
{Melis} C.,  {Dufour} P.,  2017, \mn@doi [\apj] {10.3847/1538-4357/834/1/1}, \href {https://ui.adsabs.harvard.edu/abs/2017ApJ...834....1M} {834, 1}

\bibitem[\protect\citeauthoryear{{Napier}, {Adams}  \& {Batygin}}{{Napier} et~al.}{2021a}]{napadabat2021a}
{Napier} K.~J.,  {Adams} F.~C.,   {Batygin} K.,  2021a, \mn@doi [\psj] {10.3847/PSJ/abe76e/53}, \href {https://ui.adsabs.harvard.edu/abs/2021PSJ.....2...53N} {2, 53}

\bibitem[\protect\citeauthoryear{{Napier}, {Adams}  \& {Batygin}}{{Napier} et~al.}{2021b}]{napadabat2021b}
{Napier} K.~J.,  {Adams} F.~C.,   {Batygin} K.,  2021b, \mn@doi [\psj] {10.3847/PSJ/ac29bb}, \href {https://ui.adsabs.harvard.edu/abs/2021PSJ.....2..217N} {2, 217}

\bibitem[\protect\citeauthoryear{{O'Brien}, {Tremblay}, {Klein}, {Melis}, {Koester}, {Buchan}, {Veras}  \& {Doyle}}{{O'Brien} et~al.}{2025}]{obrienetal25-1}
{O'Brien} M.~W.,  {Tremblay} P.-E.,  {Klein} B.~L.,  {Melis} C.,  {Koester} D.,  {Buchan} A.~M.,  {Veras} D.,   {Doyle} A.~E.,  2025, \mn@doi [MNRAS] {10.1093/mnras/staf398}, \href {https://ui.adsabs.harvard.edu/abs/2025MNRAS.539..171O} {539, 171}

\bibitem[\protect\citeauthoryear{{Oliverira}, {Beland}, {Keyes}, {Aloisi}, {Niemi}, {Osterman}  \& {Proffitt}}{{Oliverira} et~al.}{2010}]{olive2010}
{Oliverira} C.,  {Beland} S.,  {Keyes} C.~T.,  {Aloisi} A.,  {Niemi} S.,  {Osterman} S.,   {Proffitt} C.,  2010, in {Deustua} S.,  {Oliveira} C.,  eds, Hubble after SM4. Preparing JWST. p.~45

\bibitem[\protect\citeauthoryear{{Paganini}, {Mumma}, {Villanueva}, {DiSanti}, {Bonev}, {Lippi}  \& {Boehnhardt}}{{Paganini} et~al.}{2012}]{paganini2012}
{Paganini} L.,  {Mumma} M.~J.,  {Villanueva} G.~L.,  {DiSanti} M.~A.,  {Bonev} B.~P.,  {Lippi} M.,   {Boehnhardt} H.,  2012, \mn@doi [\apjl] {10.1088/2041-8205/748/1/L13}, \href {https://ui.adsabs.harvard.edu/abs/2012ApJ...748L..13P} {748, L13}

\bibitem[\protect\citeauthoryear{{Piani}, {Marrocchi}, {Rigaudier}, {Vacher}, {Thomassin}  \& {Marty}}{{Piani} et~al.}{2020}]{pianietal20-1}
{Piani} L.,  {Marrocchi} Y.,  {Rigaudier} T.,  {Vacher} L.~G.,  {Thomassin} D.,   {Marty} B.,  2020, \mn@doi [Science] {10.1126/science.aba1948}, \href {https://ui.adsabs.harvard.edu/abs/2020Sci...369.1110P} {369, 1110}

\bibitem[\protect\citeauthoryear{{Puzia}, {Rahatgaonkar}, {Carvajal}, {Nayak}  \& {Luco}}{{Puzia} et~al.}{2025}]{puzia2025}
{Puzia} T.~H.,  {Rahatgaonkar} R.,  {Carvajal} J.~P.,  {Nayak} P.~K.,   {Luco} B.,  2025, \mn@doi [arXiv e-prints] {10.48550/arXiv.2508.02777}, \href {https://ui.adsabs.harvard.edu/abs/2025arXiv250802777P} {p. arXiv:2508.02777}

\bibitem[\protect\citeauthoryear{{Raddi}, {G{\"a}nsicke}, {Koester}, {Farihi}, {Hermes}, {Scaringi}, {Breedt}  \& {Girven}}{{Raddi} et~al.}{2015}]{raddi2015}
{Raddi} R.,  {G{\"a}nsicke} B.~T.,  {Koester} D.,  {Farihi} J.,  {Hermes} J.~J.,  {Scaringi} S.,  {Breedt} E.,   {Girven} J.,  2015, \mn@doi [\mnras] {10.1093/mnras/stv701}, \href {https://ui.adsabs.harvard.edu/abs/2015MNRAS.450.2083R} {450, 2083}

\bibitem[\protect\citeauthoryear{{Rees}}{{Rees}}{1988}]{Rees1988}
{Rees} M.~J.,  1988, \mn@doi [\nat] {10.1038/333523a0}, \href {https://ui.adsabs.harvard.edu/abs/1988Natur.333..523R} {333, 523}

\bibitem[\protect\citeauthoryear{{Rodet} \& {Lai}}{{Rodet} \& {Lai}}{2024}]{rodlai2024}
{Rodet} L.,  {Lai} D.,  2024, \mn@doi [\mnras] {10.1093/mnras/stad3905}, \href {https://ui.adsabs.harvard.edu/abs/2024MNRAS.52711664R} {527, 11664}

\bibitem[\protect\citeauthoryear{{Rogers}, {Dorn}, {Aditya Raj}, {Schlichting}  \& {Young}}{{Rogers} et~al.}{2025}]{Rogers2025}
{Rogers} J.~G.,  {Dorn} C.,  {Aditya Raj} V.,  {Schlichting} H.~E.,   {Young} E.~D.,  2025, \mn@doi [\apj] {10.3847/1538-4357/ad9f61}, \href {https://ui.adsabs.harvard.edu/abs/2025ApJ...979...79R} {979, 79}

\bibitem[\protect\citeauthoryear{{Rubin} et~al.,}{{Rubin} et~al.}{2019}]{rubin2019}
{Rubin} M.,  et~al., 2019, \mn@doi [\mnras] {10.1093/mnras/stz2086}, \href {https://ui.adsabs.harvard.edu/abs/2019MNRAS.489..594R} {489, 594}

\bibitem[\protect\citeauthoryear{{Sahu} et~al.,}{{Sahu} et~al.}{2023}]{sahu2023}
{Sahu} S.,  et~al., 2023, \mn@doi [\mnras] {10.1093/mnras/stad2663}, \href {https://ui.adsabs.harvard.edu/abs/2023MNRAS.526.5800S} {526, 5800}

\bibitem[\protect\citeauthoryear{{Schatzman}}{{Schatzman}}{1945}]{Evry1945}
{Schatzman} E.,  1945, Annales d'Astrophysique, \href {https://ui.adsabs.harvard.edu/abs/1945AnAp....8..143S} {8, 143}

\bibitem[\protect\citeauthoryear{{Seligman} et~al.,}{{Seligman} et~al.}{2022}]{seligman2022}
{Seligman} D.~Z.,  et~al., 2022, \mn@doi [\psj] {10.3847/PSJ/ac75b5}, \href {https://ui.adsabs.harvard.edu/abs/2022PSJ.....3..150S} {3, 150}

\bibitem[\protect\citeauthoryear{{Stone}, {Metzger}  \& {Loeb}}{{Stone} et~al.}{2015}]{stoneetal15-1}
{Stone} N.,  {Metzger} B.~D.,   {Loeb} A.,  2015, \mn@doi [MNRAS] {10.1093/mnras/stu2718}, \href {https://ui.adsabs.harvard.edu/abs/2015MNRAS.448..188S} {448, 188}

\bibitem[\protect\citeauthoryear{{Str{\o}m} et~al.,}{{Str{\o}m} et~al.}{2020}]{strom2020}
{Str{\o}m} P.~A.,  et~al., 2020, \mn@doi [\pasp] {10.1088/1538-3873/aba6a0}, \href {https://ui.adsabs.harvard.edu/abs/2020PASP..132j1001S} {132, 101001}

\bibitem[\protect\citeauthoryear{{Trierweiler}, {Doyle}, {Melis}, {Walsh}  \& {Young}}{{Trierweiler} et~al.}{2022}]{trier2022}
{Trierweiler} I.~L.,  {Doyle} A.~E.,  {Melis} C.,  {Walsh} K.~J.,   {Young} E.~D.,  2022, \mn@doi [\apj] {10.3847/1538-4357/ac86d5}, \href {https://ui.adsabs.harvard.edu/abs/2022ApJ...936...30T} {936, 30}

\bibitem[\protect\citeauthoryear{{Vazan}, {Sari}  \& {Kessel}}{{Vazan} et~al.}{2022}]{vazan2022}
{Vazan} A.,  {Sari} R.,   {Kessel} R.,  2022, \mn@doi [\apj] {10.3847/1538-4357/ac458c}, \href {https://ui.adsabs.harvard.edu/abs/2022ApJ...926..150V} {926, 150}

\bibitem[\protect\citeauthoryear{{Veras} \& {Heng}}{{Veras} \& {Heng}}{2020}]{verhen2020}
{Veras} D.,  {Heng} K.,  2020, \mn@doi [\mnras] {10.1093/mnras/staa1632}, \href {https://ui.adsabs.harvard.edu/abs/2020MNRAS.496.2292V} {496, 2292}

\bibitem[\protect\citeauthoryear{{Veras}, {Shannon}  \& {G{\"a}nsicke}}{{Veras} et~al.}{2014a}]{verasetal14-3}
{Veras} D.,  {Shannon} A.,   {G{\"a}nsicke} B.~T.,  2014a, \mn@doi [MNRAS] {10.1093/mnras/stu2026}, \href {https://ui.adsabs.harvard.edu/abs/2014MNRAS.445.4175V} {445, 4175}

\bibitem[\protect\citeauthoryear{{Veras}, {Leinhardt}, {Bonsor}  \& {G{\"a}nsicke}}{{Veras} et~al.}{2014b}]{veras2014}
{Veras} D.,  {Leinhardt} Z.~M.,  {Bonsor} A.,   {G{\"a}nsicke} B.~T.,  2014b, \mn@doi [\mnras] {10.1093/mnras/stu1871}, \href {https://ui.adsabs.harvard.edu/abs/2014MNRAS.445.2244V} {445, 2244}

\bibitem[\protect\citeauthoryear{{Veras}, {Eggl}  \& {G{\"a}nsicke}}{{Veras} et~al.}{2015}]{VEG2015}
{Veras} D.,  {Eggl} S.,   {G{\"a}nsicke} B.~T.,  2015, \mn@doi [\mnras] {10.1093/mnras/stv1417}, \href {https://ui.adsabs.harvard.edu/abs/2015MNRAS.452.1945V} {452, 1945}

\bibitem[\protect\citeauthoryear{{Vernet} et~al.,}{{Vernet} et~al.}{2011}]{vernetetal11-1}
{Vernet} J.,  et~al., 2011, \mn@doi [A\&A] {10.1051/0004-6361/201117752}, \href {https://ui.adsabs.harvard.edu/abs/2011A&A...536A.105V} {536, A105}

\bibitem[\protect\citeauthoryear{{Walters}, {Farihi}, {Marsh}, {Breedt}, {Cauley}, {von Hippel}  \& {Hermes}}{{Walters} et~al.}{2023}]{Walters2023}
{Walters} N.,  {Farihi} J.,  {Marsh} T.~R.,  {Breedt} E.,  {Cauley} P.~W.,  {von Hippel} T.,   {Hermes} J.~J.,  2023, \mn@doi [\mnras] {10.1093/mnras/stac3603}, \href {https://ui.adsabs.harvard.edu/abs/2023MNRAS.519.1381W} {519, 1381}

\bibitem[\protect\citeauthoryear{{Welsh} \& {Montgomery}}{{Welsh} \& {Montgomery}}{2018}]{welsh2018}
{Welsh} B.~Y.,  {Montgomery} S.~L.,  2018, \mn@doi [\mnras] {10.1093/mnras/stx2800}, \href {https://ui.adsabs.harvard.edu/abs/2018MNRAS.474.1515W} {474, 1515}

\bibitem[\protect\citeauthoryear{{Williams}, {G{\"a}nsicke}, {Swan}, {O'Brien}, {Izquierdo}, {Cutolo}  \& {Cunningham}}{{Williams} et~al.}{2024}]{jamie2024}
{Williams} J.~T.,  {G{\"a}nsicke} B.~T.,  {Swan} A.,  {O'Brien} M.~W.,  {Izquierdo} P.,  {Cutolo} A.~M.,   {Cunningham} T.,  2024, \mn@doi [\aap] {10.1051/0004-6361/202450509}, \href {https://ui.adsabs.harvard.edu/abs/2024A&A...691A.352W} {691, A352}

\bibitem[\protect\citeauthoryear{{Williams} et~al.,}{{Williams} et~al.}{2025}]{Williams2025}
{Williams} J.~T.,  et~al., 2025, \mn@doi [\mnras] {10.1093/mnras/staf1034}, \href {https://ui.adsabs.harvard.edu/abs/2025MNRAS.541.1377W} {541, 1377}

\bibitem[\protect\citeauthoryear{{Wilson}, {G{\"a}nsicke}, {Farihi}  \& {Koester}}{{Wilson} et~al.}{2016}]{wilson2016}
{Wilson} D.~J.,  {G{\"a}nsicke} B.~T.,  {Farihi} J.,   {Koester} D.,  2016, \mn@doi [\mnras] {10.1093/mnras/stw844}, \href {https://ui.adsabs.harvard.edu/abs/2016MNRAS.459.3282W} {459, 3282}

\bibitem[\protect\citeauthoryear{{Wilson}, {Farihi}, {G{\"a}nsicke}  \& {Swan}}{{Wilson} et~al.}{2019}]{wilson2019}
{Wilson} T.~G.,  {Farihi} J.,  {G{\"a}nsicke} B.~T.,   {Swan} A.,  2019, \mn@doi [\mnras] {10.1093/mnras/stz1050}, \href {https://ui.adsabs.harvard.edu/abs/2019MNRAS.487..133W} {487, 133}

\bibitem[\protect\citeauthoryear{{Wyatt}, {Bonsor}, {Jackson}, {Marino}  \& {Shannon}}{{Wyatt} et~al.}{2017}]{wyaetal2017}
{Wyatt} M.~C.,  {Bonsor} A.,  {Jackson} A.~P.,  {Marino} S.,   {Shannon} A.,  2017, \mn@doi [\mnras] {10.1093/mnras/stw2633}, \href {https://ui.adsabs.harvard.edu/abs/2017MNRAS.464.3385W} {464, 3385}

\bibitem[\protect\citeauthoryear{{Xu}, {Zuckerman}, {Dufour}, {Young}, {Klein}  \& {Jura}}{{Xu} et~al.}{2017}]{xu2017}
{Xu} S.,  {Zuckerman} B.,  {Dufour} P.,  {Young} E.~D.,  {Klein} B.,   {Jura} M.,  2017, \mn@doi [\apjl] {10.3847/2041-8213/836/1/L7}, \href {https://ui.adsabs.harvard.edu/abs/2017ApJ...836L...7X} {836, L7}

\bibitem[\protect\citeauthoryear{{Young}, {Shahar}  \& {Schlichting}}{{Young} et~al.}{2023}]{young2023}
{Young} E.~D.,  {Shahar} A.,   {Schlichting} H.~E.,  2023, \mn@doi [\nat] {10.1038/s41586-023-05823-0}, \href {https://ui.adsabs.harvard.edu/abs/2023Natur.616..306Y} {616, 306}

\bibitem[\protect\citeauthoryear{{Zhou}, {Liu}  \& {Lin}}{{Zhou} et~al.}{2024}]{zhou2024}
{Zhou} W.-H.,  {Liu} S.-F.,   {Lin} D. N.~C.,  2024, \mn@doi [\aap] {10.1051/0004-6361/202449271}, \href {https://ui.adsabs.harvard.edu/abs/2024A&A...687A.107Z} {687, A107}

\bibitem[\protect\citeauthoryear{{Zuckerman}, {Koester}, {Reid}  \& {H{\"u}nsch}}{{Zuckerman} et~al.}{2003}]{zuckerman2003}
{Zuckerman} B.,  {Koester} D.,  {Reid} I.~N.,   {H{\"u}nsch} M.,  2003, \mn@doi [\apj] {10.1086/377492}, \href {https://ui.adsabs.harvard.edu/abs/2003ApJ...596..477Z} {596, 477}

\bibitem[\protect\citeauthoryear{{Zuckerman}, {Koester}, {Melis}, {Hansen}  \& {Jura}}{{Zuckerman} et~al.}{2007}]{zuckerman2007}
{Zuckerman} B.,  {Koester} D.,  {Melis} C.,  {Hansen} B.~M.,   {Jura} M.,  2007, \mn@doi [\apj] {10.1086/522223}, \href {https://ui.adsabs.harvard.edu/abs/2007ApJ...671..872Z} {671, 872}

\makeatother
\end{thebibliography}

\section*{Acknowledgements}
We thank the anonymous referee for valuable comments. This research is based on observations made with the NASA/ESA Hubble Space Telescope obtained from the Space Telescope Science Institute, which is operated by the Association of Universities for Research in Astronomy, Inc., under NASA contract NAS 5–26555. These observations are associated with program 12474 and 12869.  This research is based on observations collected at the European Organisation for Astronomical Research in the Southern Hemisphere under ESO programme 115.28GM.001. SS and BTG received funding from the European Research Council under the European Union’s Horizon 2020 research and innovation programme number 101002408 (MOS100PC) and 101020057 (WDPLANETS). BTG acknowledges support from the Fundaci\'on Occident and the Instituto de Astrof\'isica de Canarias under the Visiting Researcher Programme 2022-2025 agreed between both institutions.

\section{Data availability}
The COS spectroscopy data underlying this paper are available in the raw form via the \textit{HST} MAST archive under the programs 12474 and 12869. The X-shooter spectra from VLT is available in ESO archive under the program 115.28GM.001.

\newpage
\appendix
\section{Analysis of the snapshot spectrum}\label{appendix}
\begin{figure}
    \centering   
    \includegraphics[width=\columnwidth]{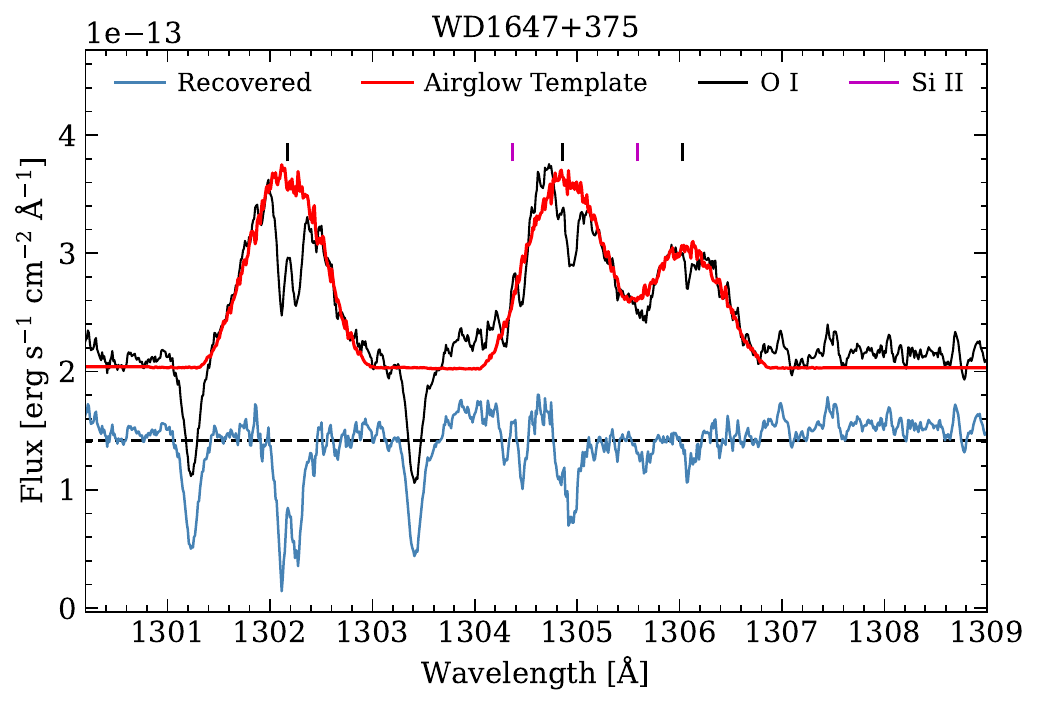}
     \includegraphics[width=\columnwidth]{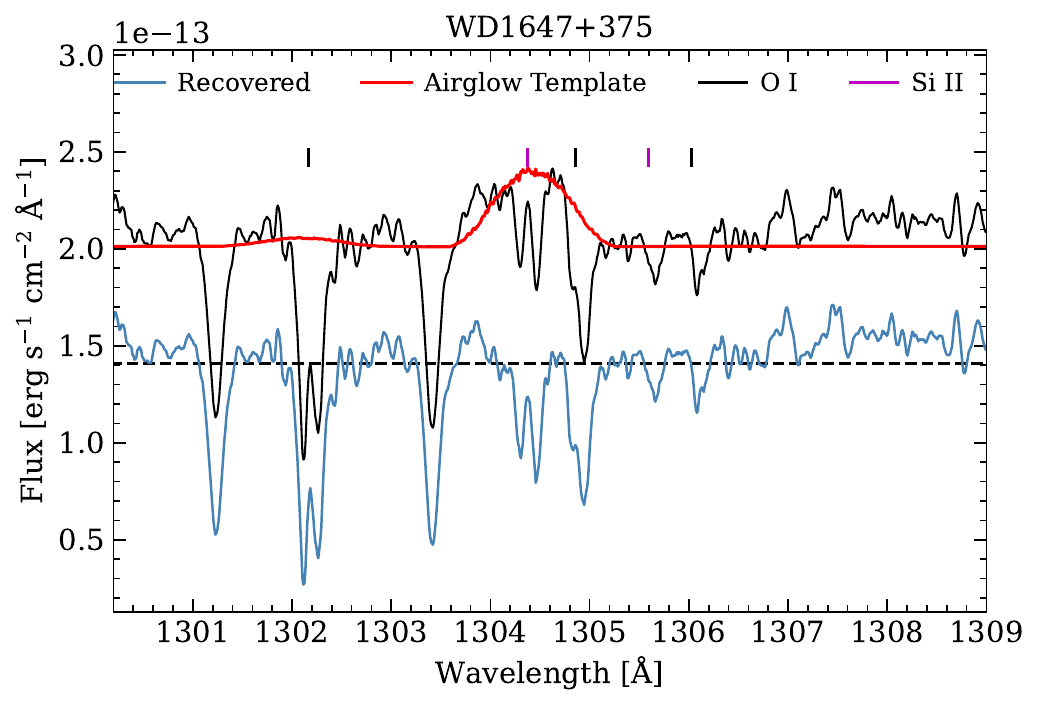}
    \caption{Airglow correction of the snapshot spectrum of WD\,1647+375 where the raw and airglow fitted template spectra are shown in solid black and red lines, respectively. The initial fit (top panel) partially removed the airglow feature in the spectral region 1304$-$1305\,\AA. Therefore, a second fit (bottom panel) was applied to the output of the first correction to get the final spectra for abundance analysis. For more description, refer Fig.\,\ref{fig:ag_corr}.}
    \label{fig:ag_corr1}
\end{figure}

\begin{figure*}
    \centering
    \includegraphics[width=\textwidth]{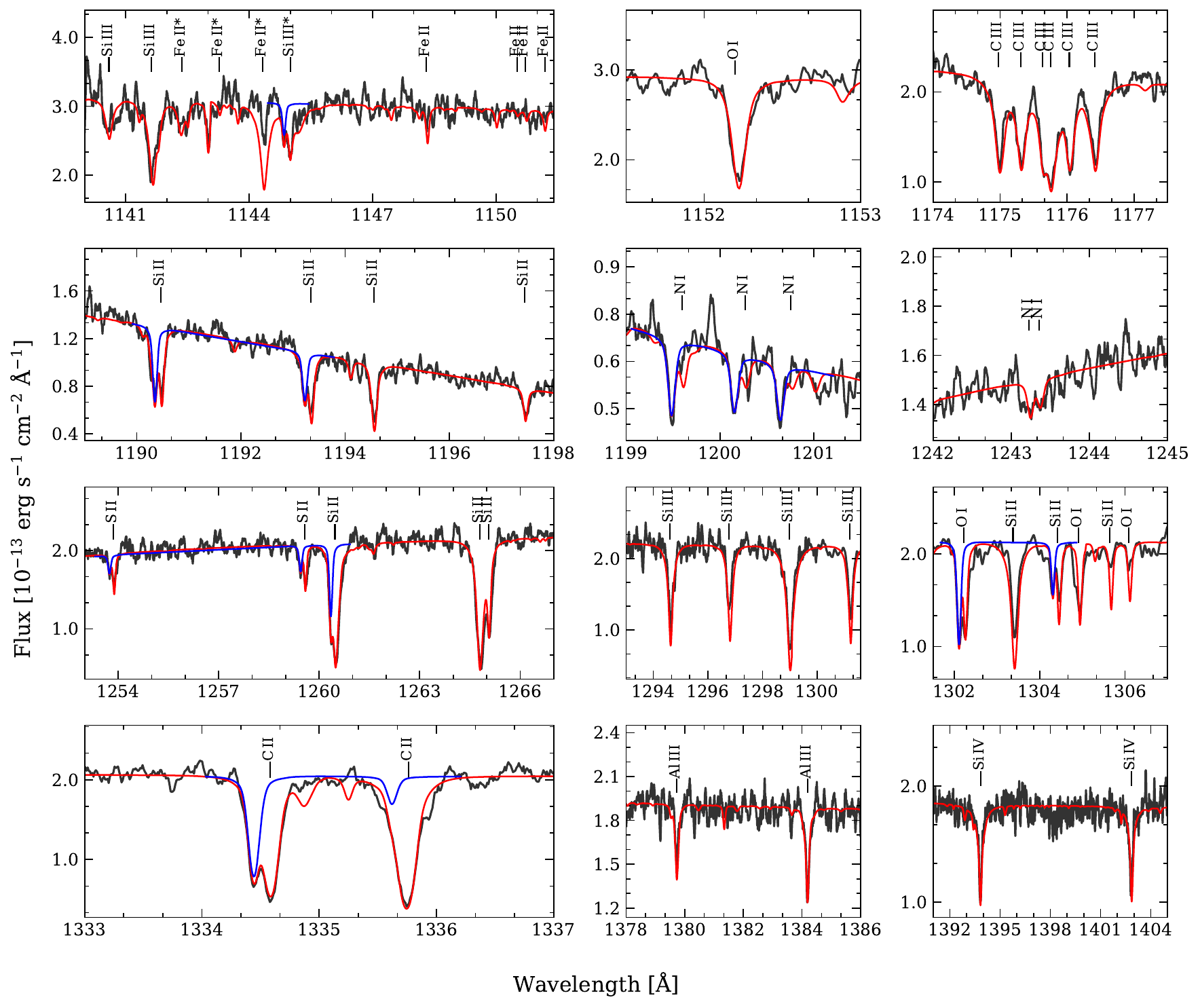}
    \caption{The airglow-corrected \textit{HST} COS snapshot spectrum of WD\,1647+375 (black) and the best-fit model (red) with the line velocities from Table\,\ref{tab:vel_comp} and the weighted average abundances from Table\,\ref{tab:abund_params}. Since we have plotted the average abundances, the individual fits to the Si lines look poor. The ISM lines are fitted with a Gaussian profile and are shown in blue. Several transitions of C, N, O, S, Si, Al, and Fe are labelled, where * denotes a blend of Si with Fe lines. }
    \label{fig:hst_spec_fit1}
\end{figure*}

The airglow correction (Fig.\,\ref{fig:ag_corr1}) and abundance analysis (Fig.\,\ref{fig:hst_spec_fit1}) for the snapshot spectra of WD\,1647+375 was performed following the same procedure as for the deeper exposure, see Sec.\,\ref{sec:analysis} in the main text for details. 

\label{lastpage}
\end{document}